\documentclass[usenatbib]{mnras}

\usepackage{newtxtext,newtxmath}

\usepackage[T1]{fontenc}
\usepackage{ae,aecompl}
\usepackage{bm}

\usepackage{graphicx}	
\usepackage{amsmath}	
\usepackage{amssymb}	

\title[MHD origin of eclipsing time variations in post-common-envelope binaries]{Magneto-hydrodynamical origin of eclipsing time variations in post-common-envelope binaries for solar mass secondaries}

\author[F. H. Navarrete et al.]{
Felipe H. Navarrete,$^{1}$\thanks{E-mail: felnavarrete@udec.cl}
Dominik R.G. Schleicher,$^{1}$
Petri~J.~K{\"a}pyl{\"a},$^{2,3}$
Jennifer Schober,$^{4}$\newauthor
Marcel V\"olschow,$^{5}$
Ronald E. Mennickent$^1$
\\
$^{1}$Departamento de Astronom\'ia, Facultad Ciencias F\'isicas y Matem\'aticas, Universidad de Concepci\'on, Av.\ Esteban Iturra\\ s/n Barrio Universitario, Casilla 160-C, Concepci\'on, Chile\\
$^{2}$Fakult\"at f\"ur Physik, Georg-August-Universit\"at G\"ottingen, Friedrich-Hund-Platz 1, 37077 G\"ottingen, Germany\\
$^{3}$ReSoLVE Centre of Excellence, Department of Computer Science, Aalto University, PO Box 15400, FI-00076 Aalto, Finland\\
$^{4}$Laboratoire d'astrophysique, Observatoire de Sauverny, CH - 1290 Versoix, Switzerland\\
$^{5}$Hamburg Observatory, Hamburg University, Gojenbergsweg 112, 21029 Hamburg, Germany
}

\date{Accepted XXX. Received YYY; in original form ZZZ}

\pubyear{2015}

\begin{document}
\label{firstpage}
\pagerange{\pageref{firstpage}--\pageref{lastpage}}
\maketitle

\begin{abstract}
Eclipsing time variations have been observed for a wide range of
binary systems, including post-common-envelope binaries. A frequently
proposed explanation, apart from the possibility of having a third
body, is the effect of magnetic activity, which may alter the internal
structure of the secondary star, particularly its quadrupole moment,
and thereby cause quasi-periodic oscillations. {Here we} present two
compressible non-ideal magneto-hydrodynamical (MHD) simulations of the
magnetic dynamo in a solar mass star, one of them with three times the
solar rotation rate (``slow rotator''), the other one with twenty
times the solar rotation rate (``rapid rotator''), to account for the
high rotational velocities in close binary systems. For the slow
rotator, we find that both the magnetic field and the stellar
quadrupole moment change in a quasi-periodic manner, leading to O-C
(observed - corrected times of the eclipse) variations of
$\sim0.025$~s. For the rapid rotator, the behavior of the magnetic
field as well as the quadrupole moment changes become considerably
more complex, due to the less coherent dynamo solution. The
resulting O-C variations are of the order $0.13$~s. The observed system
V471~Tau shows two modes of eclipsing time variations, with amplitudes
of $151$~s and $20$~s, respectively. {However, the current
  simulations may not capture all relevant effects due to the neglect
  of the centrifugal force and self-gravity. Considering the model
  limitations and that the rotation of V471~Tau is still a
  factor of 2.5 faster than our rapid rotator, it may be
  conceivable to reach the observed magnitudes.} 

\end{abstract}




\section{Introduction}

Post-common-envelope binaries (PCEBs) are close binaries that consist of a low-mass main-sequence star and a white dwarf (WD). Such systems are expected to form when the more massive component evolves until its surface extends beyond the outer Lagrangian point and eventually engulfs its companion \citep{Paczynski76}. Then the less massive star experiences friction and deposits orbital energy and angular momentum into the common envelope (CE). It spirals inwards until the envelope is expelled due to the energy transfer, leaving a close binary, typically consisting of an M dwarf (dM) or a subdwarf and a white dwarf \citep[see e.g.,][]{Parsons13}. The CE model has been revised and extended by various authors, including \citet{Meyer79, Iben93, Taam00, Webbink08} and \citet{Taam10}.

For about $90\%$ of these systems, eclipsing time variations have been
observed \citep[e.g.][]{Zorotovic13, Bours2016}. The variations occur
on rather long timescales of the order of $20$ years or more. Two
possible interpretations of these variations are commonly discussed in
the literature: The first is the presence of a third body, i.e. a
planet or a brown dwarf, which would cause apparent eclipsing time
variations due to the light-travel time (LTT) effect, i.e. the change
in the light travel time to the observer due to the change of distance
as the PCEB rotates around the common center of mass
\citep[e.g.][]{Beuermann10, Beuermann13}. Clearly, this effect
requires rather massive planets ($>M_J$, where $M_J$ is the mass of
Jupiter) on extended orbits ($>$AU) to
produce significant variations. Alternatively, such variations may
also be produced through the Applegate mechanism \citep{applegate92},
which will be described in further detail below.

If the LTT is adopted, the eclipsing time variations imply the
presence of two planets with masses of $5.6\ M_J$ and $2.1\ M_J$ and
semi-major axis of $5$~AU and $3.4$~AU, respectively, in the system NN
Ser \citep{Beuermann10, Beuermann13} which often serves as a
reference system for typical PCEBs. \citet{Beuermann13} demonstrated
the dynamical stability of these orbits, which was independently
confirmed by \citet{Horner12}. However, when the additional data by
\citet{Bours2016} are considered, they require an {extra} quadratic
term in the expression for the eclispsing times if they want to maintain the planet solution. The physical
origin of such an additional term is however unclear. In case of the
system HW~Vir, a two-planet solution appears to be secularly stable
\citep{Beuermann12}. A final conclusion on the stability of orbits in
Hu~Aqr is still pending \citep{Godziewski12, Hinse12, Bours2014,
  Godz2015}, and similarly for QS~Vir \citep{Parsons10b}. In the case of
the system V471~Tau, the proposed third body has been searched via
Direct Imaging, but has not been found \citep{Hardy2015}. Using the
orbital period of the system (12.5-hr) and the spin period (9.25-min)
of the WD as two independent clocks, \citet{Vanderbosch2017} have
concluded that a third body interpretation cannot adequately explain
the nature of this system.

If the planets in  NN~Ser are real, they should also be dynamically young, as the white dwarf has an age of only $10^6$~yrs \citep{Parsons10a}. While their existence is highly speculative, they have at least two possible origins. The so-called first-generation scenario proposes that they formed together with the binary and then survived the common envelope phase, while the second-generation scenario implies that they formed through the material ejected during the common envelope phase.  A hybrid scenario may be also possible, with accretion of the ejected gas onto already existing planets. Several studies have been carried out on this matter \citep[e.g.,][]{Volschow13, Schleicher14, Bear14}, though it is currently difficult to draw any final conclusions. 

The other possible explanation of the eclipsing time variations is
magnetic activity. Historically, in particular the Applegate mechanism
\citep{applegate92} has been a relevant scenario, in which the
magnetic activity of the secondary stars leads to a redistribution of
the stellar angular momentum, thus changing its gravitational
quadrupole moment. This in turn produces a variation of the binary
separation. The original Applegate model has been improved by several
authors. For instance, \citet{Lanza99} improved the model by adopting
a consistent description of stellar virial
equilibrium. \citet{Brinkworth06} extended the model introducing a
finite shell formalism, considering the exchange of the angular
momentum between the shell and the core. \citet{Volschow16} examined
their model in more detailed and applied it to a sample of $16$ close
binary systems (predominantly PCEBs), showing that the Applegate
mechanism is a viable process in the shortest and most massive binary
systems. The corresponding model has been made public through the
Applegate calculator\footnote{Applegate calculator:
  \url{http://theorygroup-concepcion.cl/applegate/index.php}}, and shows
that the mechanism is favored in particular for rapidly rotating
systems \citep{Navarrete2018}.

In addition to the finite shell model, \citet{Lanza2004} and
\citet{Lanza2005} presented a one-dimensional framework based on the
angular momentum transport equations, using simplifying assumptions of
magneto-hydrodynamical (MHD) turbulence and the mean magnetic
field. We have extended this framework in \citet{Volschow2018},
considering in particular time-dependent hydrodynamic and magnetic
fluctuations assuming a magnetic activity cycle, as well as a
superposition of different modes. For typical RS Canum Venaticorum (RS~CVn) systems, which
are detached binaries typically composed of a chromospherically active
G or K star, the expected eclipsing time variations are however two
orders of magnitude lower than observed. The most promising Applegate
candidates are post-common-envelope binaries with secondary masses of
$\sim0.35$~M$_\odot$ \citep{Volschow2018}, as these produce more
energy through nuclear burning and can thus more easily redistribute
angular momentum as required by the Applegate mechanism, while simultaneously not being critically
affected by the presence of a radiative core.

The presence of magnetic activity should be expected in these systems
due to the convective envelopes of the secondaries and their rapid
rotation. A corresponding dynamo model has been
put forward by \citet{Ruediger02}. Observationally, magnetic activity
has been inferred on many occasions. In the case of V471 Tau, it has
been probed via photometric variability, flaring events and H$\alpha$
emission along with a strong X-ray signal \citep{Kaminski07,
  Pandey08}. For DP~Leo, magnetic activity has been revealed through
X-ray observations \citep{Schwope02}. In the system QS~Vir, it is
indicated via detections of Ca~II emission and Doppler imaging
\citep{Ribeiro10}, as well as observed coronal emission
\citep{Matranga12}. In case of HR~1099, a 40 year X-ray light curve
suggesting a long-term cycle was recently compiled by
\citet{Perdelwitz2018}, and similar studies have been pursued via
optical data \citep[e.g.,][]{Donati2003, Lanza2006, Berdyugina2007,
  Muneer2010}.

While magnetic activity is potentially relevant to explain the origin
of the eclipsing time variations, its effects on the stellar structure
so far have only been explored via finite shell or 1D models, in both
of which the presence of a dynamo was externally imposed. However, a
self-consistent modeling of the dynamo and its interaction with the
stellar structure may be crucial, and is only possible within 3D
magneto-hydrodynamical (MHD) simulations. While stellar dynamo models have
previously been pursued \citep[see e.g.][]{Yadav2016}, the latter
was done in the anelastic limit, which does not allow to
explore the effect of the dynamo onto the stellar structure. Here as a
first step, we will employ a fully compressible setup developed by
\citet{kapyla13} for
a solar mass star although with rotation rates exceeding the solar
one. These models allow the quantification of changes in stellar
structure due to the dynamo. While solar mass stars are not
very common in post-common-envelope systems, they do occasionally
occur, as for instance the secondary of V471~Tau has a mass of $0.93\pm0.07$~M$_\odot$
\citep{Zorotovic13}, and is thus still consistent with being a solar
mass star. Independently, we of course stress that this is an
exploratory study that should be extended to stars of different masses
as well.

In section~\ref{pencil}, we will briefly introduce the {\sc Pencil
  Code}\footnote{\url{https://github.com/pencil-code/}}
\citep{Brandenburg02, Brandenburg03} as well as the setup employed
here, which is based on previous developments by \citet{kapyla13}. In
these simulations, the Rayleigh number, which describes the ratio of the time scale for thermal transport via diffusion to the time scale for thermal transport via convection, is however
much smaller than in reality due to the higher diffusivities required
for numerical stability, see detailed discussions in \cite{kapyla13}
and \cite{KupMut17}. Another caveat of fully compressible simulations
of solar-like stars is that the low Mach number in the deep parts of
the convection zone necessitates a very short time step and that the
thermal relaxation occurs in the Kelvin--Helmholtz timescale which is
of the order of $10^7$ {($10^5$) years for the whole Sun (solar
convection zone)} \citep[see,
  e.g.][]{KupMut17}. Thus, to bring the dynamic and acoustic
timescales closer to each other and to shorten the Kelvin--Helmholtz
timescale, the energy flux needs to be enhanced \citep[see
  also][]{Brandenburg05}. To compensate for this and to
obtain a comparable rotational influence on the flow as in real stars,
which is the key factor determining their dynamo properties, the
angular velocity needs to be increased proportional to one third power
of the increase of the energy flux \citep[see a detailed description
  in][]{KGOKB19}. For this reason, the effect of the centrifugal force
has
been omitted in this formulation of the Navier-Stokes equation, as the
resulting centrifugal force would be too high, thereby significantly
altering the hydrostatic balance \citep{kapyla11,kapyla13}. With this
in mind, we note that our simulations present only a first step, where
the redistribution of material can be explored for instance due to
meridional flows, and we will not probe the effect originally proposed
by \citet{applegate92}. Nevertheless, the occurence of quadrupole
moment variations even in the absence of the centrifugal force term is
a central outcome of the simulations. The results of our simulations
are presented in section~\ref{results}, including a hydrodynamical
reference run and two MHD simulations. Our discussion and conclusions
are presented in section~\ref{conclusions}.

\section{Methods}\label{pencil}

\subsection{Pencil Code}

The \textsc{Pencil code}
\citep{Brandenburg02, Brandenburg03} is a finite-difference code
written in Fortran 95. It uses sixth-order spatial derivatives
and a third-order Runge-Kutta time integrator scheme, which makes the
code particularly useful for studying weakly compressible turbulent
flows. For the timestepping, a high-order scheme is implemented in
order to reduce amplitude errors and to allow longer time steps, which
is the \textit{RK-2N} Runge-Kutta scheme \citep{williamson80}, where
the ``2N" stands for its memory consumption of two chunks. The time step is specified by the Courant time step. The \textit{Message passing interface} (MPI) is used for parallelization.

\subsection{The model}\label{subsec:pencilsetup}

The model we use here is based on {those used by} \citet{Cole2014}
and \citet{viviani18} and is {described}
here for completeness. The computational domain is spherical but
without the poles, which allows to reach a higher spatial resolution
but at the cost of omitting connecting flows across the poles and
introducing artificial boundaries at high latitudes. The domain
$(r,\theta,\phi)$ denotes radial, colatitudinal, and longitudinal
directions. The radius extends from $0.7~R_\odot$ (the bottom of the
convection zone) to $1.0~R_\odot$, {where $R_\odot$ is the solar
  radius;} $\theta$ goes from $\pi/12$ to
$11\pi/12$, and $\phi$ from $0$ to $2\pi$. The corresponding grid resolution is $128\times256\times512$. We employ the compressible
non-ideal MHD equations in the following form:
\begin{align}
 \frac{\partial \boldsymbol{A}}{\partial t} &= \boldsymbol{u}\times\boldsymbol{B} - \mu_0\eta\boldsymbol{J}, \label{eq:induction} \\
 \frac{D\ln\rho}{Dt} &= - \bm\nabla\cdot\boldsymbol{u}, \label{eq:conserv-mass} \\
 \frac{D\boldsymbol{u}}{Dt} &= \boldsymbol{g} - 2\boldsymbol{\Omega}_0\times\boldsymbol{u}+\frac{1}{\rho}\left(\boldsymbol{J}\times\boldsymbol{B}-\bm\nabla p+\bm\nabla\cdot2\nu\rho\bm{\mathsf{S}}\right), \label{eq:conserv-momentum} \\
 T\frac{Ds}{Dt} &= \frac{1}{\rho}\left\{-\bm\nabla\cdot\left(\boldsymbol{F}^\textnormal{rad}+\boldsymbol{F}^\textnormal{SGS}\right)+\mu_0\eta\boldsymbol{J}^2\right\}+2\nu\bm{\mathsf{S}}^2, \label{eq:conserv-energy}
\end{align}
where $\boldsymbol{A}$ is the magnetic vector potential,
$\boldsymbol{u}$ and $\boldsymbol{B} = \bm\nabla\times\boldsymbol{A}$ are
the velocity and magnetic
field, $\boldsymbol{J} = \mu_0^{-1}\bm\nabla\times\boldsymbol{B}$ is
the electric current density with $\mu_0$ being the vacuum permeability. $D/Dt
= \partial/\partial t+\boldsymbol{u}\cdot\bm\nabla$
is the convective derivative, $\rho$ is the density, and
\begin{equation}
 \boldsymbol{F}^\textnormal{rad} = -K\bm\nabla T,
\end{equation}
 and
 \begin{equation}
 \boldsymbol{F}^\textnormal{SGS} = -\chi_\textnormal{SGS}\rho T\bm\nabla s,
\end{equation}
are the radiative and subgrid scale (SGS) fluxes. The former accounts
for the flux coming from the radiative core and the latter is added to
stabilize the scheme and to account for the unresolved turbulent
transport of heat. $K$
and $\chi_\textnormal{SGS}$ are the radiative heat conductivity and
and turbulent entropy diffusivity, respectively. $s$ is the specific
entropy, $p$ is the pressure,
and $T$ is temperature. Furthermore, the system of equations
(\ref{eq:induction})--(\ref{eq:conserv-energy}) is closed by assuming an
ideal gas law,
\begin{equation}
 p = (\gamma - 1)\rho e, \label{eq:idealgas}
\end{equation}
where $\gamma = c_P/c_V = 5/3$ is the ratio of specific heats at
constant pressure and volume, and $e = c_V T$ is the specific internal
energy. $\bm{\mathsf{S}}$ is the traceless rate-of-strain tensor
\begin{equation}\label{eq:rateofstrain}
  \mathsf{S}_{ij} = \frac{1}{2}(u_{i;j}+u_{j;i}) - \frac{1}{3}\delta_{ij}\bm\nabla\cdot\bm{u},
\end{equation}
where semicolons denote covariant differentiation. $\bm{g} =
-GM\hat{\bm r}/r^2$ is the gravitational acceleration with $G$ being the
gravitational constant, $M$ the stellar mass, and $\hat{\bm r}$ the radial
unit vector. The stellar rotation vector is given as
$\mathbf{\Omega}_0 = (\cos\theta, -\sin\theta,0)\Omega_0$. As already
discussed in the introduction, the formulation of the Navier-Stokes
equation employed here does not include the centrifugal force term,
which would be unrealistically high \citep[see][for
  details]{kapyla11,kapyla13,KGOKB19}.

\subsection{Initial and boundary conditions}

The initial state is isentropic with a temperature gradient given as
\begin{equation}
 \frac{\partial T}{\partial r} = -\frac{GM/r^2}{c_V(\gamma -1)(n_\textnormal{ad}+1)},
\end{equation}
where $n_{\rm ad}=3/2$ is the polytropic index for adiabatic
stratification.
The fixed values that define a simulation are (i) the energy flux at
the bottom,
\begin{equation}
 F_b = -K\left(\frac{\partial T}{\partial r}\right)\Big|_{r = r_0},
\end{equation}
where $K = (n+1)K_0$ is the radiative conductivity, $K_0$ is a
constant \citep{kapyla13}, and
\begin{equation}
 n = 2.5\left(\frac{r}{r_0}\right)^{-15} - 1.
\end{equation}
Here $n = n_{\rm ad}$ at the bottom and $n \rightarrow -1$ at the
surface. This choice is made to ensure that the radiative flux at the
bottom is solely responsible for supplying energy into the system and
that convection transport essentially the total flux in the bulk of
the convection zone (CZ). The remaining parameters of the model are
(ii) the angular velocity $\Omega_0$, (iii) viscosity $\nu$, (iv)
magnetic diffusivity $\eta$, and (v) turbulent heat conductivity
$\chi_\textnormal{SGS}$ and its radial
profile \citep[see][]{kapyla13}. The {turbulent} velocity and magnetic fields are initialized with
small-scale low amplitude Gaussian noise perturbations.

\subsubsection{Radial boundary}

The radial boundaries are assumed to be impenetrable and stress-free, i.e at $r = r_0,\,R$:
\begin{align}
 u_r &= 0, \\
 \frac{\partial u_\theta}{\partial r} &= \frac{u_\theta}{r}, \\
 \frac{\partial u_\phi}{\partial r} &= \frac{u_\phi}{r}.
\end{align}
The bottom $(r = r_0 = 0.7R)$ is assumed to be a perfect conductor with
\begin{equation}
 \frac{\partial A_r}{\partial r} = A_\theta = A_\phi = 0,
\end{equation}
and at the top $(r = R)$ the magnetic field is radial
\begin{align}
 A_r &= 0, \\
 \frac{\partial A_\theta}{\partial r} &= -\frac{A_\theta}{r}, \\
 \frac{\partial A_\phi}{\partial r} &= -\frac{A_\phi}{r}.
\end{align}
The value of $\partial T / \partial r$ is fixed at the bottom and the
upper radial boundary uses a black body condition
\begin{equation}
 \sigma T^4 = -K\nabla_rT - \chi_\textnormal{SGS}\rho T \nabla_r s,
\end{equation}
where $\sigma$ is a modified value of the Stefan-Bolzmann constant \citep[see][]{kapyla13}.

\subsubsection{Latitudinal boundary}

The latitudinal boundary is also assumed to be stress-free at $\theta = 15$\textdegree, 165\textdegree
\begin{align}
 \frac{\partial u_r}{\partial\theta} &= u_\theta = 0, \\
 \frac{\partial u_\phi}{\partial \theta} &= u_\phi\cot\theta,
\end{align}
and a perfect conductor
\begin{equation}
 A_r = \frac{\partial A_\theta}{\partial \theta} = A_\phi = 0.
\end{equation}
Density and entropy are assumed to have zero first derivative on both boundaries, thus suppressing heat fluxes through them.

\subsection{Quadrupole moment and its scaling}\label{rescaling}

The quadrupole tensor is defined as
\begin{equation}
Q_{ij} = I_{ij} - {1 \over 3} \delta_{ij} {\rm Tr}~I,
\end{equation}
where ${\rm Tr}$ denotes the trace and $I_{ij}$ is the {tensor of inertia}
\begin{equation}
I_{ij} = \int x_i x_j dm = \int \rho({\bm x}) x_i x_j d^3x,
\end{equation}
where $x_i$ refer to Cartesian coordinates.

As already mentioned in the introduction, the stellar luminosity is
enhanced due to numerical constraints \citep[see][]{Brandenburg05,
KGOKB19}. As a result, the energy flux coming from the
bottom is much higher than in the Sun. The ratio of fluxes
$\mathfrak{F}_r$ in the present case is
\begin{equation}
  \mathfrak{F}_r = \frac{\mathfrak{F}_{\rm simulation}}{\mathfrak{F}_\odot} = 8.07 \cdot 10^5.
\end{equation}
The increased flux implies that the fluctuations of other quantities
are correspondingly enhanced. The fluctuation of the pressure can be
written as
\begin{equation}
  \Delta p = \left(\frac{\partial p}{\partial \rho}\right)_s \Delta \rho \equiv c_{\rm s}^2 \Delta \rho, \label{equ:delp1}
\end{equation}
where the subscript $s$ indicates constant entropy and where $c_s$ is
the sound speed. Furthermore, variations in pressure scale as
\begin{equation}
\Delta p \sim \rho \boldsymbol{u}^2.\label{equ:delp2}
\end{equation}
Equating (\ref{equ:delp1}) and (\ref{equ:delp2}) we obtain
\begin{equation}\label{eq:mach}
 \frac{\Delta\rho}{\rho} \sim \frac{\boldsymbol{u}^2}{c_s^2} = \rm{Ma}^2.
\end{equation}
Here Ma is the Mach number, which scales as \citep[e.g.][]{KGOKB19,K19}
\begin{equation}
 \rm{Ma} \sim \mathfrak{F}_r^{1/3},
\end{equation}
and thus,
\begin{equation}
 \Delta \rho \sim \mathfrak{F}_r^{2/3}.
\end{equation}
All of the numbers given in sections~\ref{subsec:3-qxx} and
\ref{subsec:20x-q} have been rescaled in this fashion, which
corresponds a factor of $(8.07\cdot10^5)^{-2/3} \approx
1.15\cdot10^{-4}$, i.e.
\begin{equation}
Q_{xx} = 1.15\cdot10^{-4} Q_{xx, \rm sim},
\end{equation}
where the subscript `sim' denotes the estimated quadrupole moment obtained
in the simulations.

{Furthermore, we define the Taylor, Coriolis, fluid and magnetic
  Reynolds, and SGS and magnetic Prandtl numbers as
\begin{gather}
 {\rm Ta} = \!\left[ \frac{2\Omega_0(0.3R)^2}{\nu}\right]^2,\ {\rm Co} = \frac{2\Omega_0}{u_{\rm rms} k_1}, \\
 {\rm Re} = \frac{u_{\rm rms}}{\nu k_1},\ {\rm Re_M} = \frac{u_{\rm rms}}{\eta k_1}, \\
 {\rm Pr_M} = \frac{\nu}{\eta},\  {\rm Pr_{\rm SGS}} = \frac{\nu}{\chi_{\rm SGS}^{\rm m}},
\end{gather}
where $u_{\rm urms}$ is the volume-averaged root-mean-square
velocity, $k_1 = 2\pi/0.3R$ is an estimate of the wavenumber of the
largest eddies, and $\chi_{\rm SGS}^{\rm m}$ is the SGS entropy
diffusion at $r=0.85~R_\odot$.}

\section{Results}\label{results}

In this section, we present our main results obtained from the
numerical simulations. In subsection~\ref{notation}, we
introduce the notation {used throughout the paper and discuss} the
overall properties of
our simulations. {We first discuss a pure hydrodynamical reference
  run in subsection \ref{hydro} to demonstrate that the long-term
  modulation of the quadrupole moment must have a
  magneto-hydrodynamical origin}. We {then} present two MHD
models, a slow rotator (3 times
solar rotation, $P_{\rm rot} = 9$~days) and a fast rotator (20 times
solar rotation, $P_{\rm rot} = 1.4$~days) in
subsections~\ref{slow} and \ref{fast}. These values were chosen as the
rotation rate in PCEBs is considerably enhanced compared to isolated
stars, with a rotation period in V471 Tau of about $0.522$~d
\citep{Zorotovic13}.


\subsection{Notation and General Properties}\label{notation}

We label the two MHD simulations according to their rotation
rates, namely \textit{\textit{run3x}} for the 3 times solar rotation, and
\textit{run20x} for the 20 times solar rotation rate. Quantities with
an overline indicate an average over the azimuthal angle;
e.g.\ $\overline{B}_r$ indicates an average of the $r$ component of $B$
over $\phi$ and is given by
\begin{equation}
  \overline{B}_r(r,\theta) = \frac{\int B_r(r,\theta,\phi)\,d\phi}{\int d\phi}.
\end{equation}
Other averages are presented inside angular brackets with sub-
and superscripts. For example, $\langle \overline{B}_r \rangle_i^k$
indicates an average of $\overline{B}_r$ in regions denoted with $i$
and $k$. The subscript indicates the depth at which the quantity of
interest is taken and the superscript indicates the latitude where the
average is further calculated with the following rules:
\begin{align}
 i &= \{\rm{s, m, b}\}, \\
 k &= \{\rm{np, eq, sp}\},
\end{align}
where
\begin{align}
{\rm s} &= {\rm surface} \rightarrow r = 0.98 R, \label{defs}\\
{\rm m} &= {\rm middle} \rightarrow r = 0.85 R, \\
{\rm b} &= {\rm bottom} \rightarrow r = 0.72 R,\label{defb}
\end{align}
and
\begin{align}
  \rm{np}& = \rm{northern\ hemisphere} & & 75^{\circ} < 90\degr - \theta < 0^{\circ}, \\
  \rm{eq} &= \rm{equator} & & 20^{\circ} < 90\degr - \theta < -20^{\circ}, \\
  \rm{sp} &= \rm{southern\ hemisphere} & & 0^{\circ} < 90\degr - \theta < -75^{\circ},
\end{align}
where $\theta$ is colatitude. So for example, $\langle \overline{B}_r
\rangle_{\rm{s}}^{\rm{eq}}$ indicates the average of the
azimuthally-averaged $B_r$ over $20^{\circ} < \theta < -20^{\circ}$,
i.e.\ the {eq}uatorial, near the {s}urface of the computational domain. 

\begin{figure}
\includegraphics[width=\linewidth]{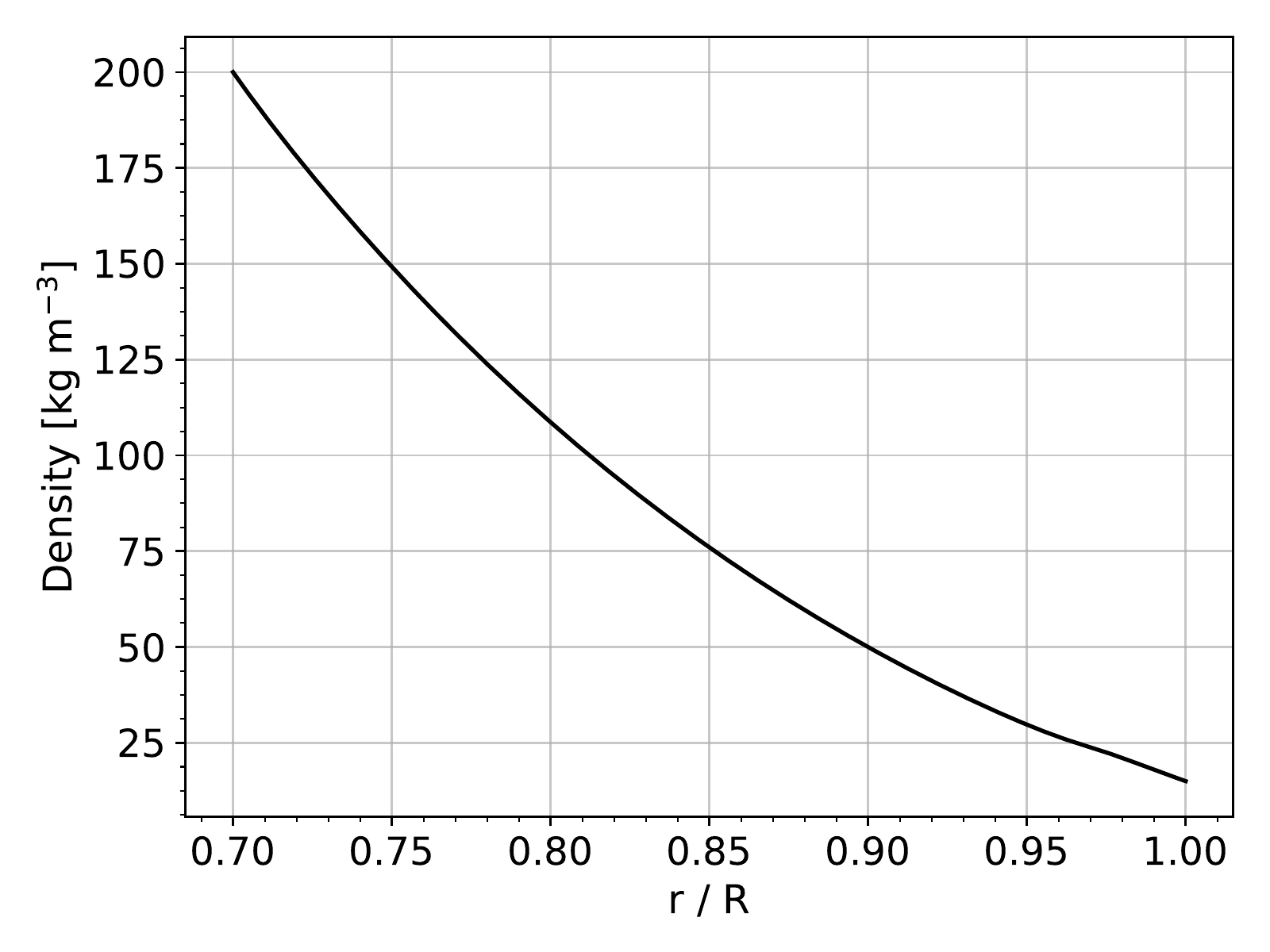}
\includegraphics[width=\linewidth]{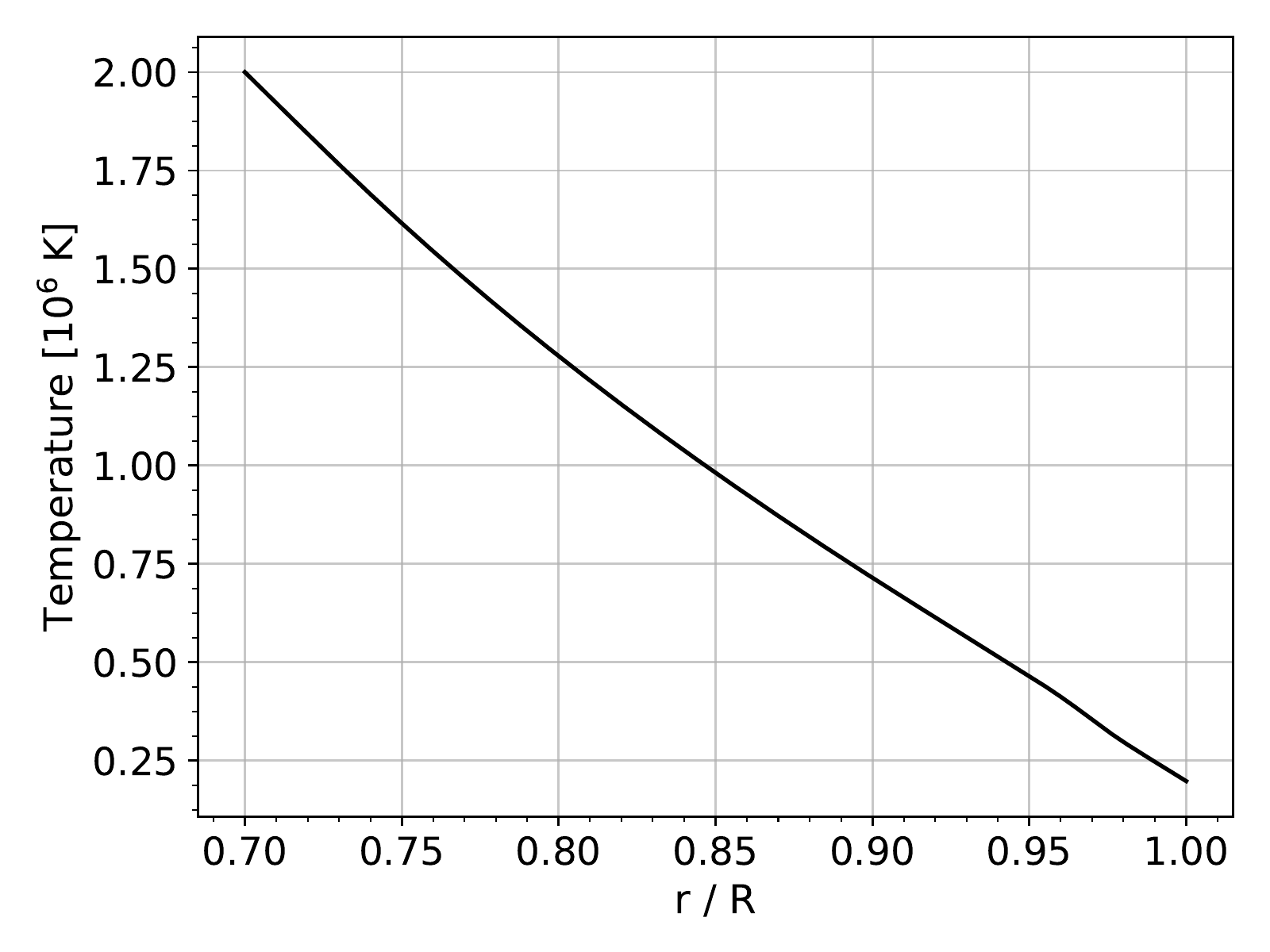}
\caption{Density (top) and temperature profile (bottom) for \textit{run20x}.}
\label{fig:20-profile-density}
\end{figure}

Typical density and temperature profiles are shown in
Figure~\ref{fig:20-profile-density},
corresponding to the final state of \textit{run20x}.
The density at the bottom is 181.8~kg~m${^{-3}}$ and
13.6~kg~m${^{-3}}$ at the surface, where bottom and surface are evaluated following the definition in (\ref{defs}) and (\ref{defb}). This corresponds to a density
\begin{equation}
 \frac{\rho_{\rm bottom}}{\rho_{\rm surface}} = 13.4.
\end{equation}
The
temperature profile is shown in the lower panel of Figure
\ref{fig:20-profile-density}. The temperature at the bottom is set
to be the same as the temperature at the bottom of the CZ
in the Sun, namely ${T =2\times10^6}$~K, and decreases towards a value
of ${1.9\times10^5}$~K at the surface.

\begin{figure}
\includegraphics[width=\linewidth]{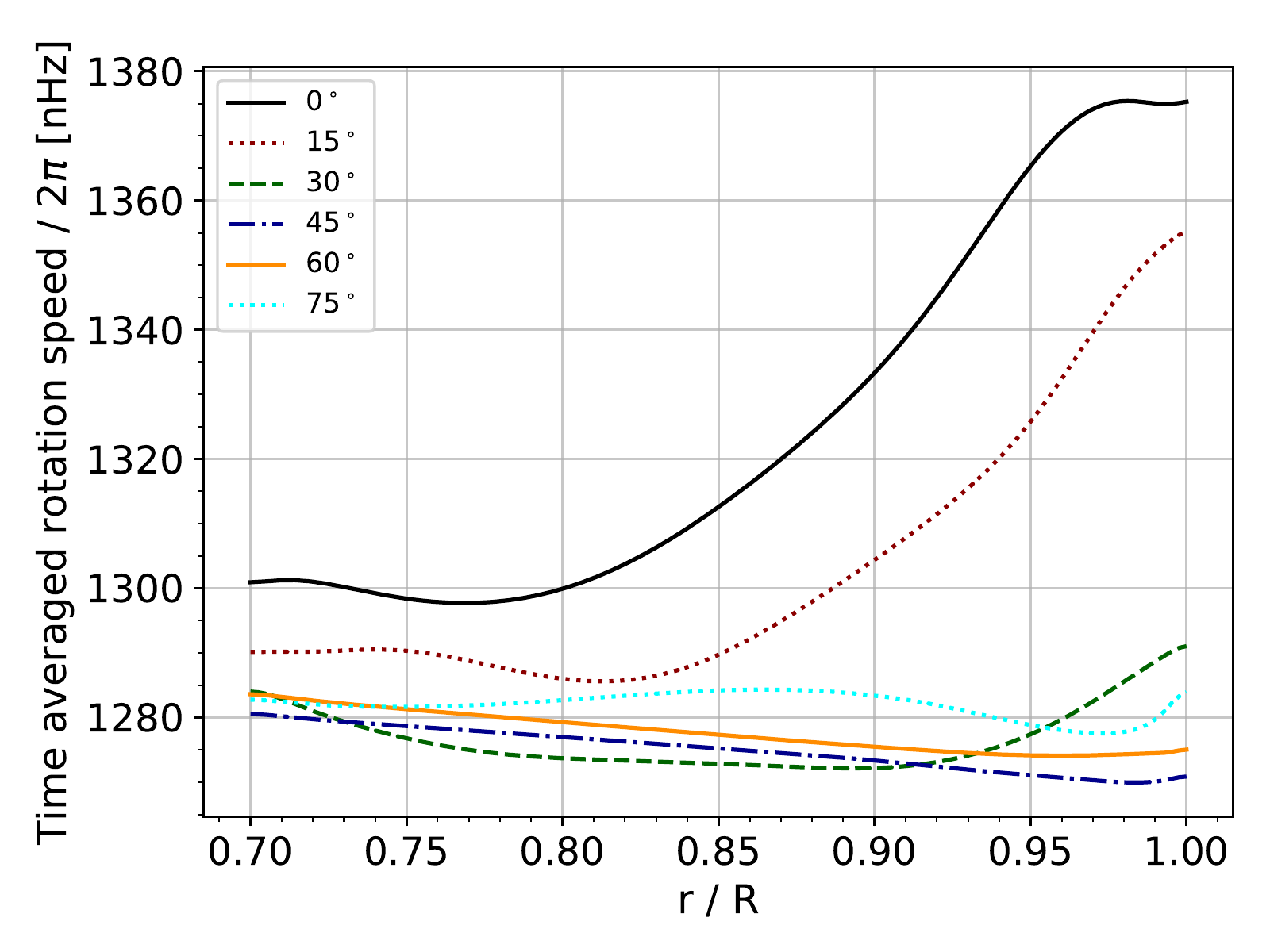}
\includegraphics[width=\linewidth]{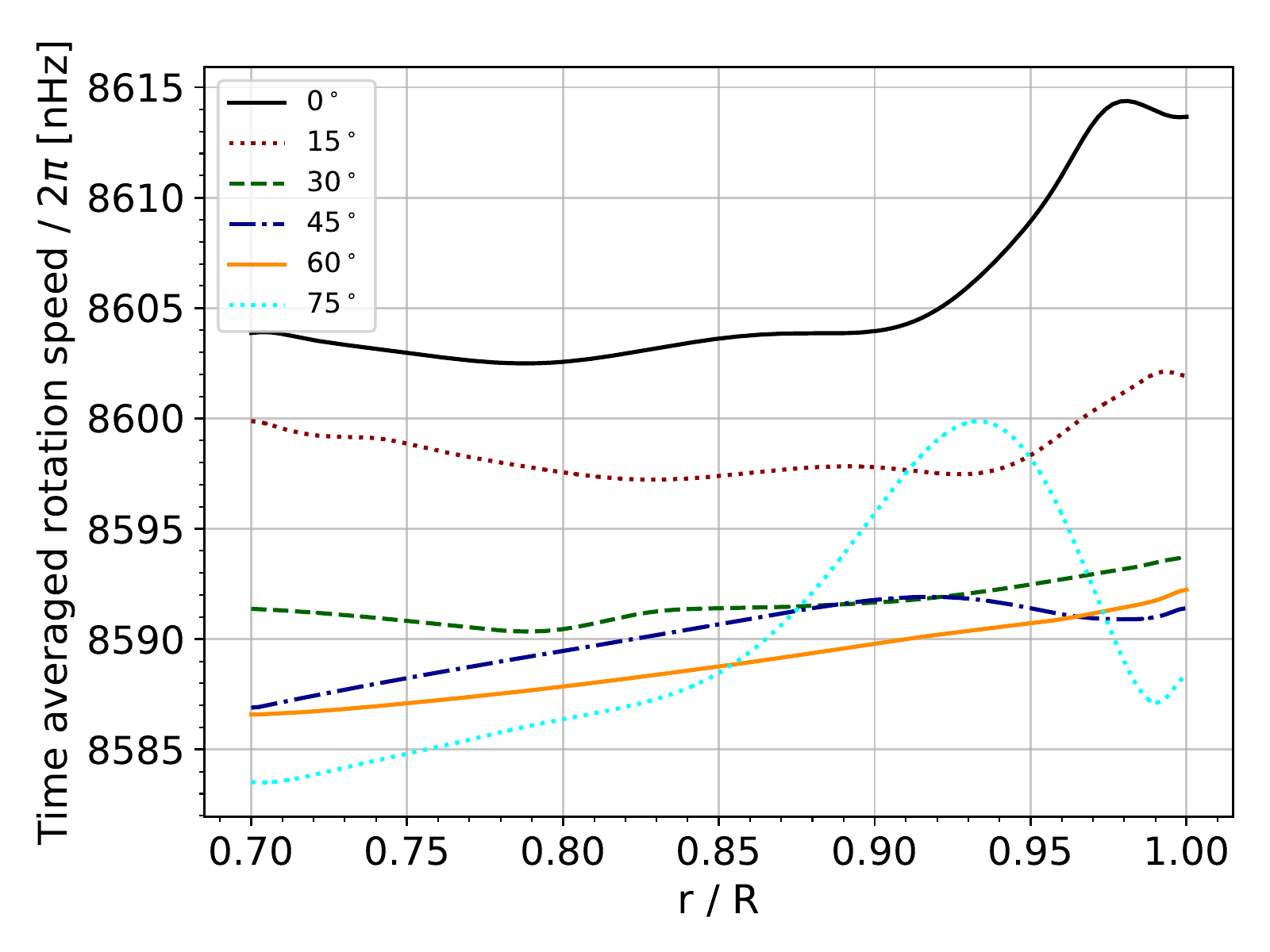}
\caption{Angular velocity as a function of radius from six latitudes
  indicated by the legend from runs \textit{run3x} (top) and \textit{run20x} (bottom).}
\label{fig:20-profile-rotation}
\end{figure}

{The time-averaged angular velocity
  $\overline{\Omega}=\overline{u}_\phi/r \sin \theta + \Omega_0$ is
  shown from six latitudes from \textit{run3x} in the top panel of
  Figure~\ref{fig:20-profile-rotation}. Overall the rotation is faster
  at the equator than at high latitudes, but we often observe an
  increase in the angular velocity at the latitude boundaries (see the
  cyan-dotted lines in Fig.~\ref{fig:20-profile-rotation}). This is
  likely an artefact due to the impenetrable latitude boundary. In the
  lower panel of Figure~\ref{fig:20-profile-rotation}, we show the
  time-averaged rotation profile for \textit{run20x}. The difference
  in the rotation rates between high latitudes and the equator is
  significantly smaller than in the slower rotator. The decrease of
  differential rotation as the overall rotation rate is increased is
  consistent with earlier studies \citep[e.g.][]{viviani18}.}

\begin{figure}
\includegraphics[width=\linewidth]{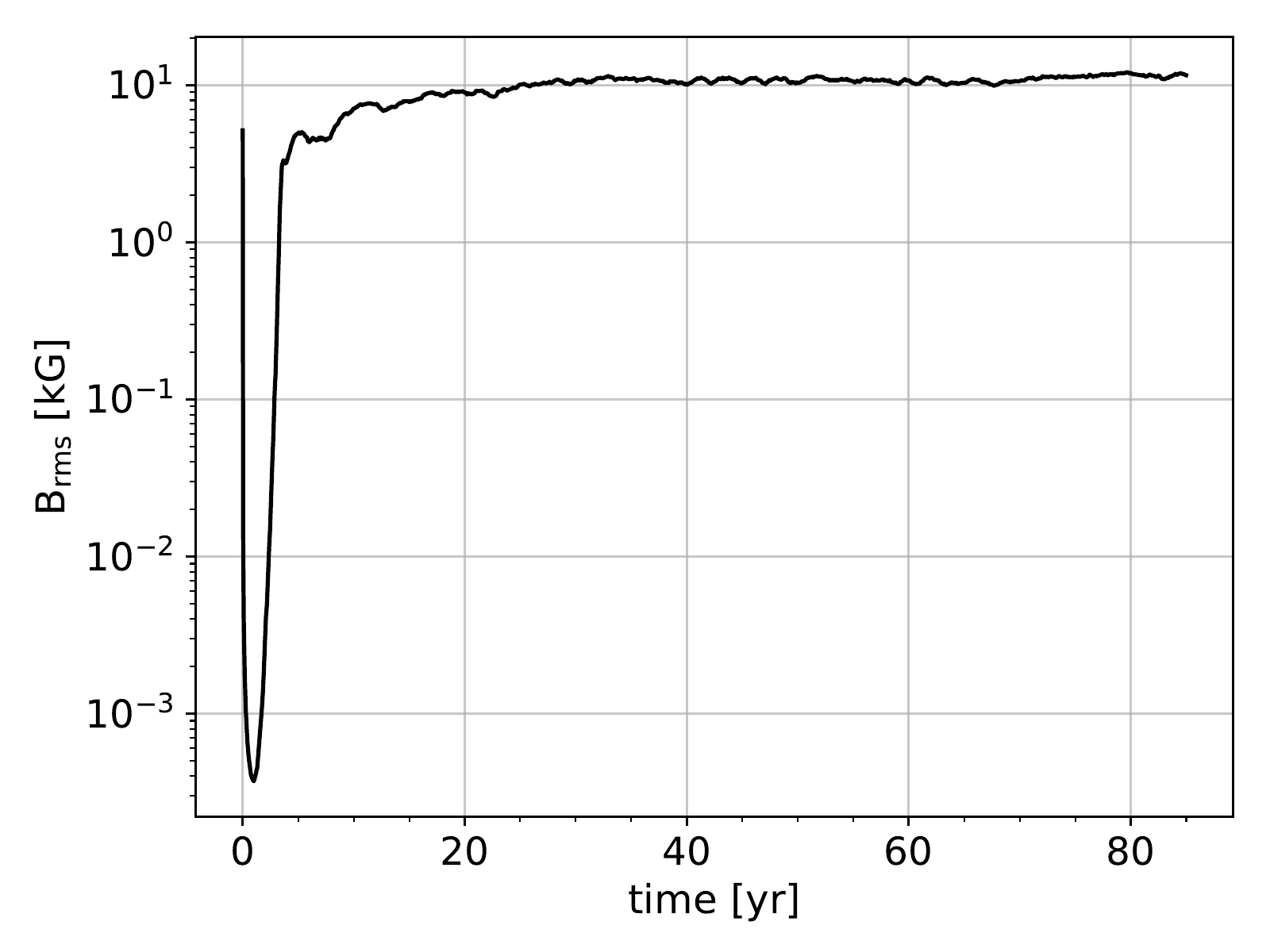}
  \caption{Saturation of the dynamo in \textit{run20x}. The rms-value of the magnetic
    field, $B_{\rm rms}$, grows exponentially up to the saturation
    regime where the analysis is performed.}
\label{fig:20-brms}
\end{figure}

At the beginning, the simulations first have to go through a
relaxation phase during which thermal and magnetic saturation is
established. The description that follows corresponds to \textit{run20x},
but is qualitatively the same for \textit{run3x}. There are two conditions that
need to be fulfilled before analyzing the results and deriving
astrophysical implications for PCEB systems:

On the one hand, the system has to reach dynamo saturation, which is
shown in Figure \ref{fig:20-brms}, where we plot the root-mean-squared
magnetic field for \textit{run20x}. The seed magnetic field first
decays because most of {the initial} magnetic energy is contained
on the small scales which is quickly dissipated \citep{dobler06}. {
  Furthermore, it takes susbtantial time for convection and
  large-scale flows to develop that lead to dynamo action.}
Subsequently, the
magnetic field grows exponentially during the next three years during
{the kinematic stage}. This growth starts to slow down in the
non-linear regime until it reaches the saturation stage after about
$22$~years.

\begin{figure}
\includegraphics[width=\linewidth]{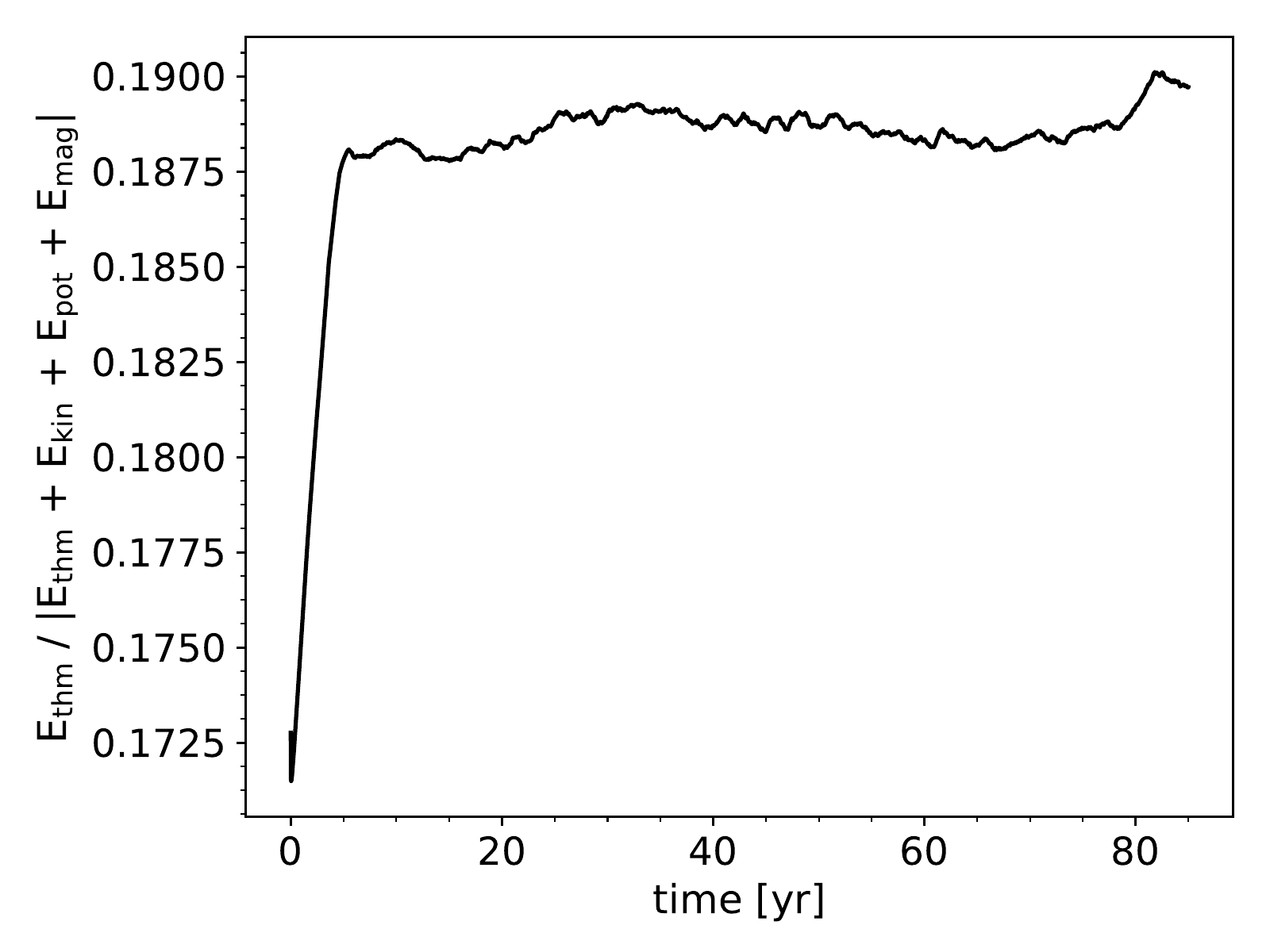}
  \caption{Evolution of the thermal energy as a fraction of the total
    energy of the system in \textit{run20x}. $E_{\rm thm}$ is the thermal energy, $E_{\rm
      kin}$ is the kinetic energy, $E_{\rm pot}$ is the potential
    energy, and $E_{\rm mag}$ is the magnetic energy. The initial
    transient is due to the onset and maturing of convection and the
    corresponding development of turbulent heat transport.}
\label{fig:20-ethm}
\end{figure}

The system has to also reach thermal {equilibrium}, which is shown in
Figure~\ref{fig:20-ethm}, where we plot the
fraction of thermal to total energy. The only energy source in the
simulations is the energy injected from the bottom {of the convection zone}.
{While most of this energy is transported to the surface by
  convection, a fraction is deposited in the thermal reservoir of the
  convection zone until equilibrium between energy input and output is
  achieved. This is manifested by an increase of the thermal energy in
  the present case. Thermal evolution after roughly 8~years is slow
  and the system is sufficiently close to thermal saturation to be
  used for statistical analysis of the data.}

%
%
%
%

\subsection{Purely hydrodynamical simulation}\label{hydro}
Here we present a pure hydrodynamical reference run with 20 times solar rotation. This serves essentially for comparison with the MHD simulations, to demonstrate that long-term variations only occur in simulations that include magnetic fields.

\begin{figure}
\includegraphics[width=\linewidth]{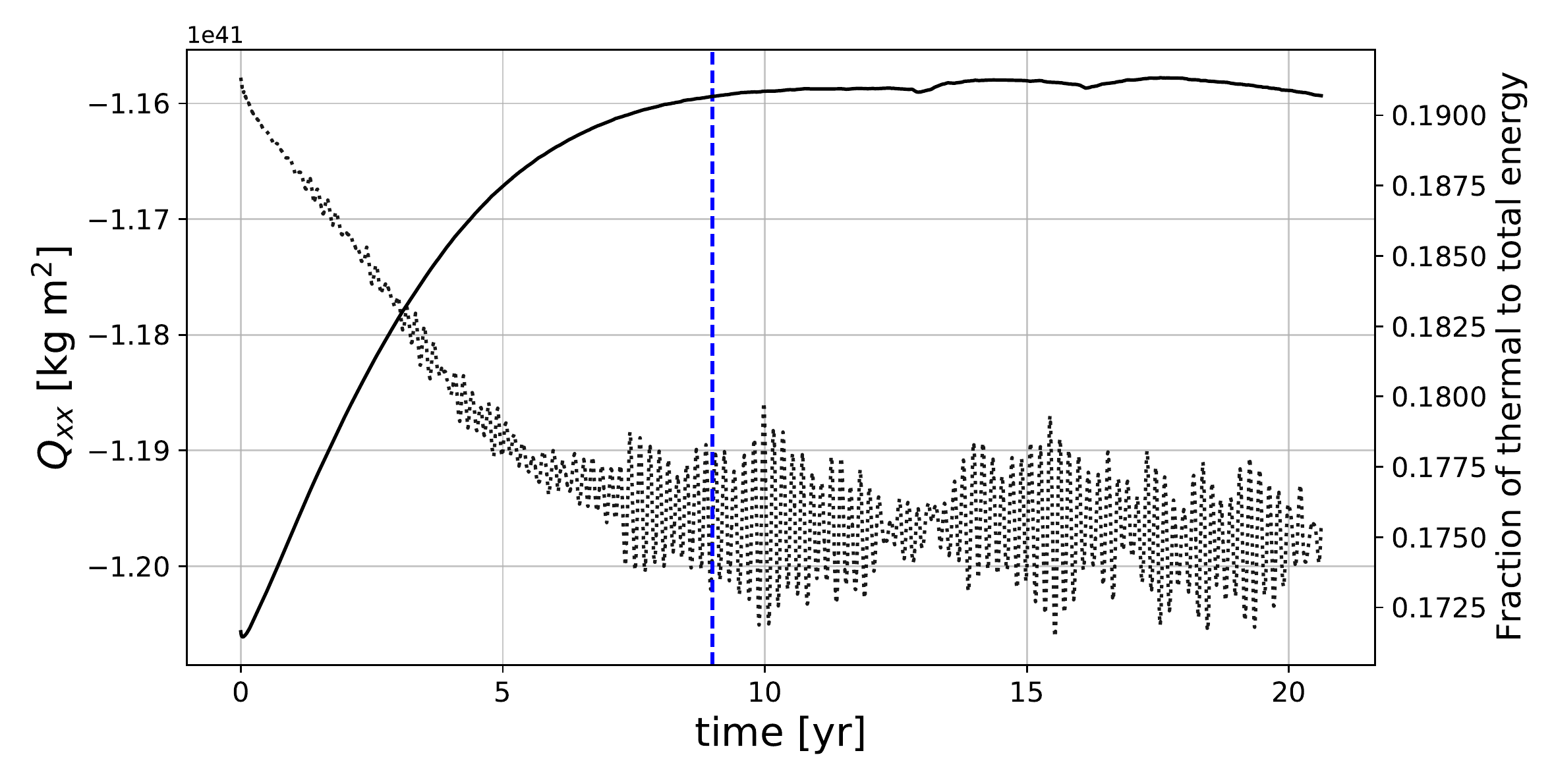}
\caption{Gravitational quadrupole moment ($Q_{xx}$; dotted line)
  variations together with the thermal energy of the system (solid
  line). High frequency oscillations are obtained. The thermal
  relaxation phase coincides with the phase of gravitational
  quadrupole moment relaxation, marked with the dashed blue line.}
\label{fig:hydro-q-ethm}
\end{figure}

Figure \ref{fig:hydro-q-ethm} shows the change of the gravitational quadrupole moment $Q_{xx}$ in the dotted line, together with the thermal energy as a fraction of the total energy of the system as solid line.
Here we can see very high frequency oscillations with a period of
0.$18$~years. This {is very close to an average} sound-crossing
time $\tau_{\rm sc}$, which we calculate as
\begin{equation}
  \tau_{\rm sc} = \frac{2\,r_{\rm conv}}{\langle c_{\rm s}\rangle_{\rm vol}} \approx 0.18\,{\rm yr},
\end{equation}
where $r_{\rm conv}$ is the radial extent of the simulations, and $\langle c_{\rm s}\rangle_{\rm vol}$ corresponds to the sound-speed, $c_{\rm s}$, averaged over the radial direction. The sound-speed is calculated as
\begin{equation}
 c_{\rm s} = \sqrt{\left(\frac{\partial p}{\partial \rho}\right)_s}
\end{equation}
where the subscript `s' indicates the derivative is taken at constant
entropy. {Thus the high frequency oscillations have a purely
hydrodynamical origin.}

%
%
%
%
%
%

\subsection{The case of the slow rotator (\textit{run3x})}\label{slow}

%
%
%
%
%
%

We first investigate the evolution {of} a simulation with three
times solar rotation. {This case is characterised by ${\rm
    Ta}=5.68\times10^7$, ${\rm Co}=2.68$, ${\rm Re}={\rm Re}_M=71$,
  ${\rm Pr}_M=1$, and ${\rm Pr}_{\rm SGS}=2.5$. Simulations with similar
  parameters were also explored by \citet{viviani18} for the stellar
  dynamo but they have not explored the implications on the stellar
  structure.}

\subsubsection{Overview of convective and magnetic states}\label{subsec:3x-surface}
\begin{figure}
\includegraphics[width=\linewidth]{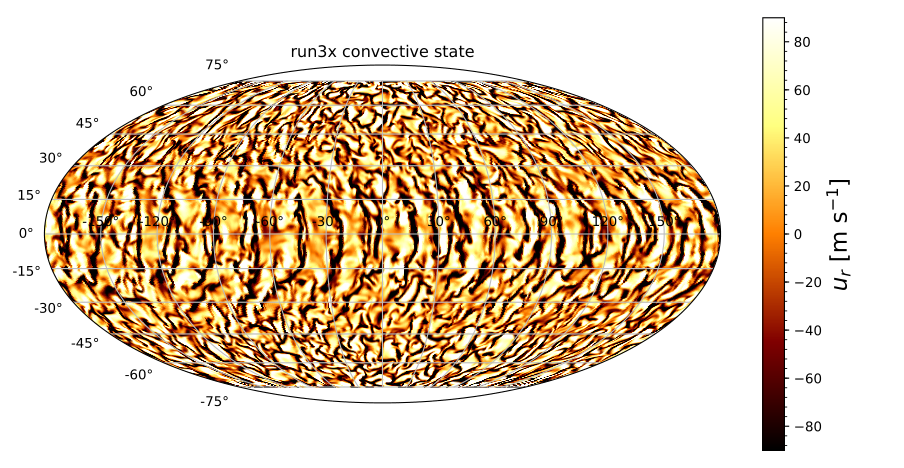}
\caption{Radial velocity near the surface for \textit{run3x}. The colorbar is cut at $\pm90$ m~s$^{-1}$ to improve visualization.}
\label{fig:3-state-convective}
\end{figure}

To illustrate the general structure, we first examine a snapshot at
the end of the simulation. Figure \ref{fig:3-state-convective} shows a
Mollweide projection (an equal-area map projection also known as
\textit{homolographic projection}) of the near surface radial velocity.
The colorbar is cut at
$\pm90$ m~s$^{-1}$ to improve visualization. The velocity field does
not show clear signs of large-scale structures.
At the equator we see elongated cells (sometimes called ``banana
cells''). Their existence is due to the influence that rotation has on
the flow \citep[see][]{Busse1970}. At higher latitudes the
the banana cells disappear, and the effect of rotation is to give rise
to more symmetric and smaller cells \citep[see, e.g.][]{Chandra61}. It
should be noted that these
cells are much larger than {those} observed in the Sun which is
due to the much lower density stratification in the simulations. The
mean
radial velocity is $\pm$60~m~s$^{-1}$, {while the extrema can reach}
$\pm$1000~m~s$^{-1}$.

\begin{figure}
\includegraphics[width=\linewidth]{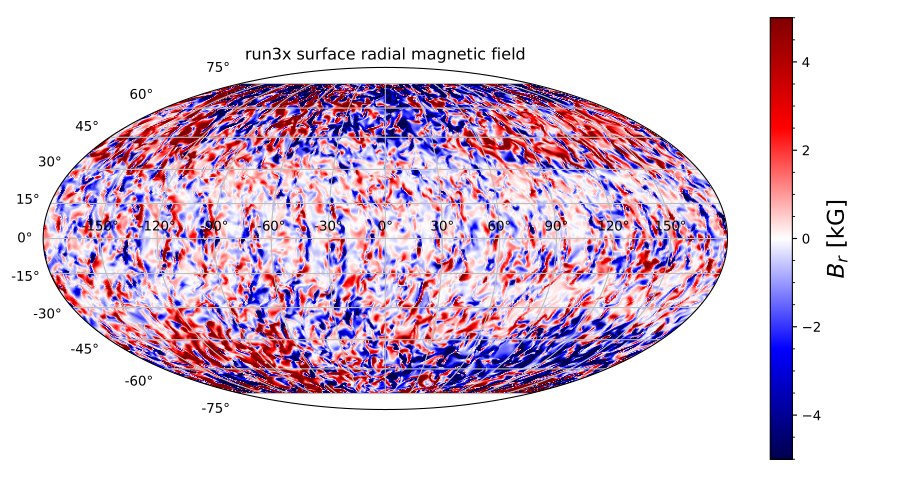}
\caption{Radial magnetic field near the surface for \textit{run3x}. The colorbar is cut at $\pm5$ kG to improve visualization.}
\label{fig:3-state-br}
\end{figure}

In Figure~\ref{fig:3-state-br} we plot the near
surface radial magnetic field at the end of the simulation. The
colorbar is cut at $\pm5$ kG to improve visualization.
The magnetic field strength at
the equator is weaker than at high latitudes, and a clear $m=1$
non-axisymmetric component is observed. The mean magnetic field
strength is 2.5~kG and the extrema are $\pm$~90~kG. The sizes of the
magnetic structures is much larger than, e.g., sunspots. This is due
to the fact that the current simulations lack the resolution to
capture the small-scale granulation near the surface and the physics
leading to spot formation.

\subsubsection{Overview of the magnetic field evolution}\label{subsec:3x-overview}

\begin{figure}
  \includegraphics[width=\linewidth]{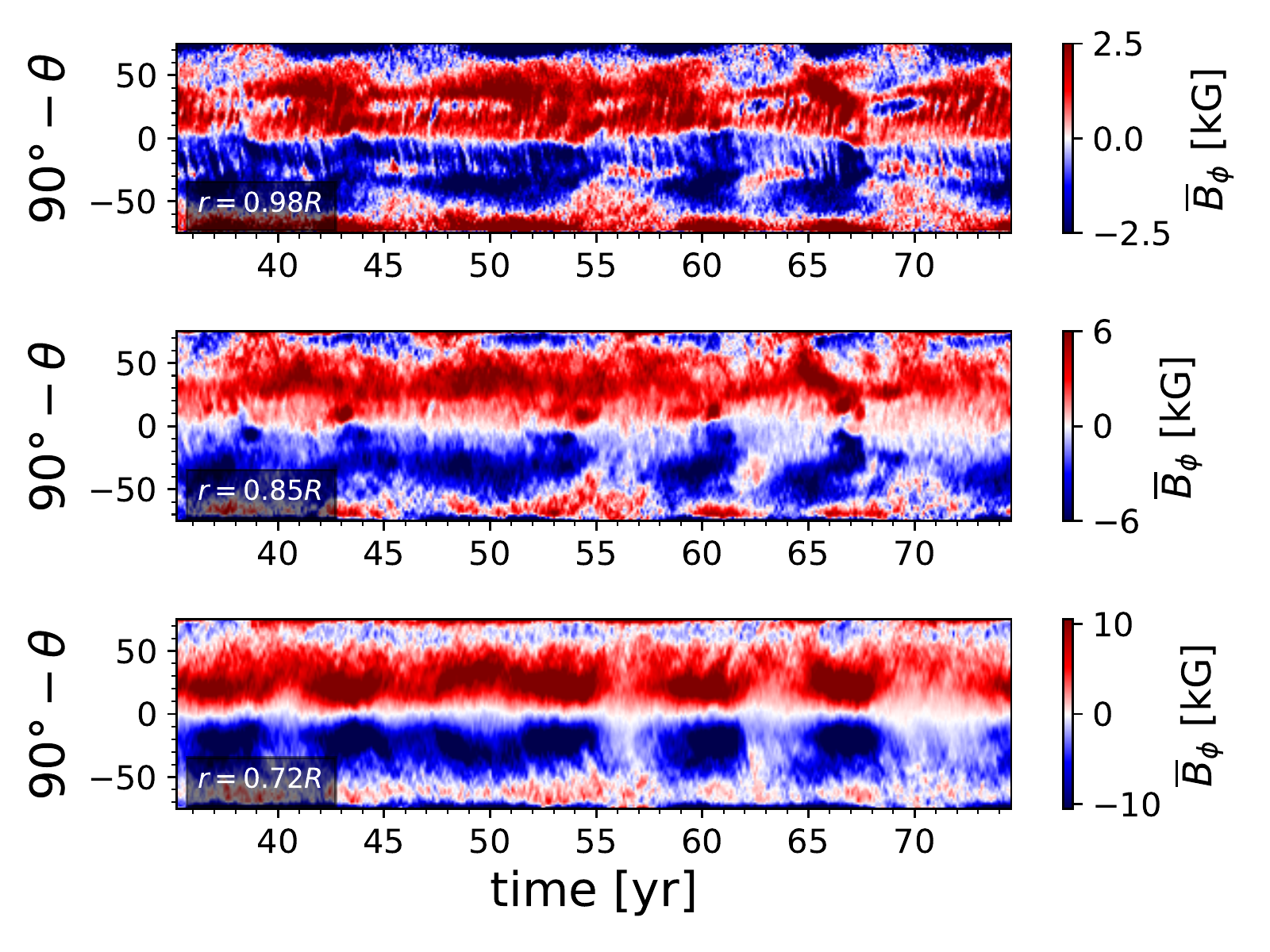}
\caption{Time evolution of the mean toroidal magnetic field
  $\overline{B}_\phi$ for \textit{run3x} at three different depths, labeled at
  the lower-left corner of each panel. The magnetic field is changing its
  intensity and there are short periods where the activity is much
  weaker. Color bars are cut to improve visualization.}
\label{fig:3-butterfly-bz}
\end{figure}

We start the analysis by examining the dynamo solution of the slow
rotator. Figure \ref{fig:3-butterfly-bz} shows the evolution of the
mean toroidal magnetic field (butterfly diagram) of \textit{run3x} at
three depths. At the northern hemisphere there
is an overall positive polarity whereas in the southern hemisphere the
dominant polarity is negative. In each
panel, the polarity near the latitude boundaries is opposite to the
dominant
polarity. Polarity reversals can be seen at high latitudes at the
bottom of the convective region (third panel). In the middle (second
panel), these reversals at the poles are more pronounced and thus
easier to see whereas at the surface the reversals are not as clearly
observed in the azimuthal field. Thus it appears that the
axisymmetric part of the magnetic field consists of a dominant
quasi-stationary component and a weaker quasi-periodic one, as also
recently reported by \citet[][]{Viviani2019}. Meanwhile, {the
  strength of the azimuthal magnetic field is varying
  quasi-periodically without polarity reversals near the equator}. At
the
three reference depths there are episodes of decreased activity, for
example at the equator during the time frames of 55 to 59 years and 62
to 66 years. The extrema at the bottom, middle, and surface are
$\pm$~20 kG, $\pm$~7 kG, and $\pm$~3 kG, respectively.

\begin{figure}
  \includegraphics[width=\linewidth]{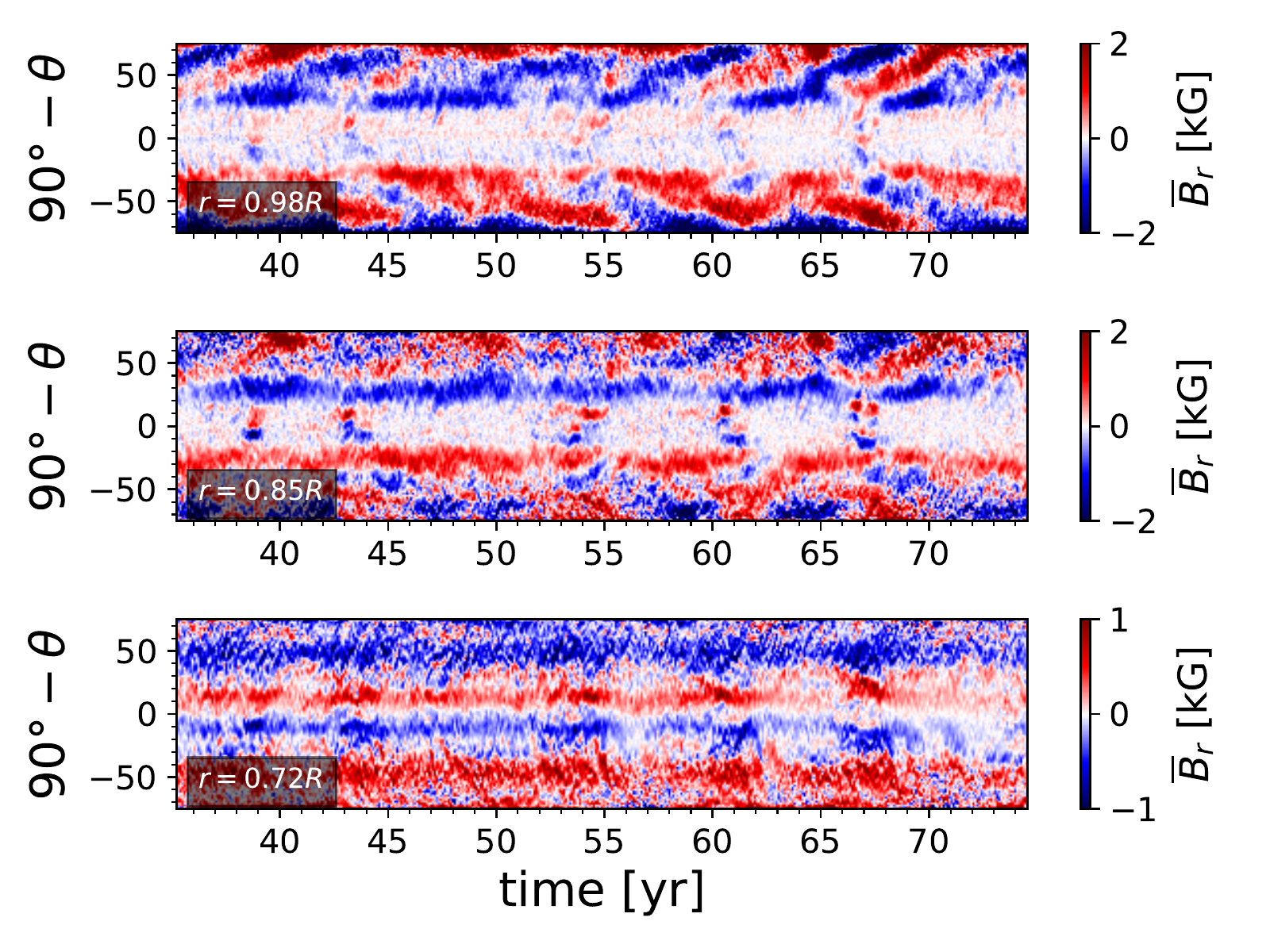}
\caption{Time evolution of the mean radial magnetic field
  $\overline{B}_r$ for \textit{run3x} at three different depths, labeled at the
  lower-left corner of each panel. A poleward migration of the
  magnetic field is clearly seen at the surface of the
  domain. Sporadic activity is seen at the equator.}
\label{fig:3-butterfly-bx}
\end{figure}

The evolution of the radial field is shown in Figure
\ref{fig:3-butterfly-bx}. At the bottom of the convective zone (bottom
panel) the behaviour of $\overline{B}_r$ is similar to the one
described for the toroidal field at the surface. At low latitudes and
towards the equator the magnetic field is positive (negative) at the
northern (southern) hemisphere, and there are no clear signs of polarity
reversals. In the middle of the convective region we start seeing
hints of a poleward migrating dynamo wave (see second panel in Figure
\ref{fig:3-butterfly-bx}) {at latitudes greater than $50^\circ$
  in both hemispheres}. Meanwhile, at mid-latitudes (${\pm 30^\circ}$)
a {persistently negative (positive) magnetic field in the northern
  (southern) hemisphere} is obtained with no
migration. Near the equator, the mean radial magnetic field is
weaker but with periods of increased strength at ${t = 39, 43, 55, 51,
  {\rm and}, 67}$~years. At the surface of the star (top panel) a
dynamo wave is obtained with a poleward migration. At the equator the
strength of $\overline{B}_r$ is weaker with periods of increased
strength at the same times as in the middle of the computational
domain.

\subsubsection{Origin of the quadrupole moment fluctuations}\label{sebsec:3-q-origin}

\begin{figure}
\includegraphics[width=\linewidth]{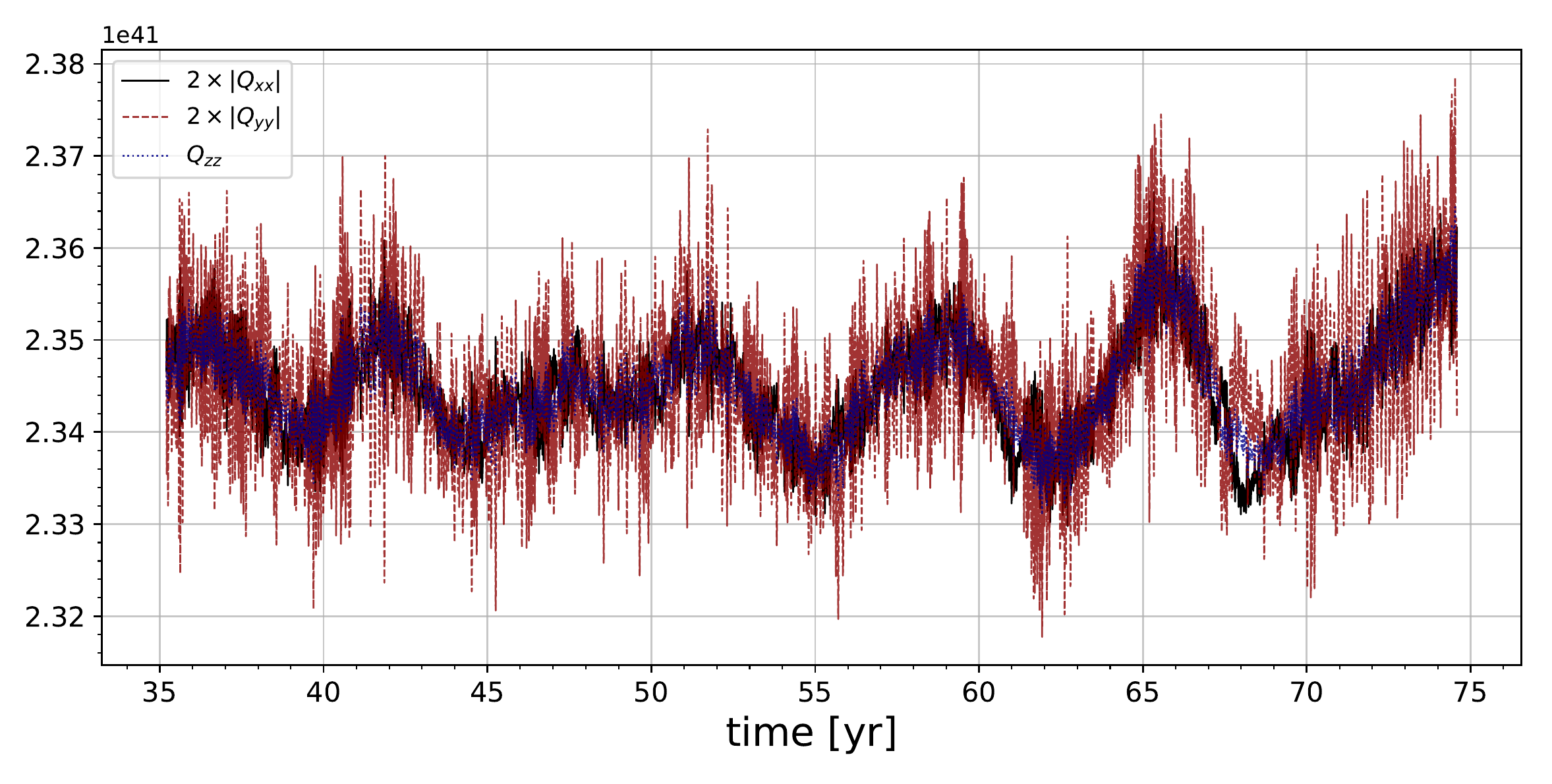}
\caption{Time evolution of the gravitational quadrupole moment
  components $2|Q_{xx}|$, $2|Q_{yy}|$ and $Q_{zz}$ in \textit{run3x}. Apart
  from differences of a factor of $\sim2$, the components follow the
  same overall trend, in the form of short-term differences in the high
  frequency oscillations.}
\label{fig:3-qall}
\end{figure}

The time evolution of {the diagonal elements of the quadrupole moment}
$Q_{xx}$, $Q_{yy}$ and
$Q_{zz}$ is shown in Fig.~\ref{fig:3-qall}. While $Q_{zz}$ is
positive, $Q_{xx}$ and $Q_{yy}$ are negative, and there is a
difference of about a factor of $2$ in the components. Apart from
that, their overall behavior is very similar, showing a quasi-periodic
evolution with a period of around 8 years, and an amplitude of the
order of $\mathbf{\sim1\times10^{39}}$~kg~m$^2$ in the case of $Q_{xx}$. For comparison, we also applied the semi-analytic model by \citet{Volschow2018}, obtaining the same order of magnitude for the fluctuations. The
system further shows the presence of short-term oscillations, which
are also present in hydrodynamical runs (see
Sect.~\ref{hydro}). We will in the following take the $Q_{xx}$
component as a reference that we compare to other quantities, keeping
in mind that the result would be similar for the other components as
well.

\begin{figure}
\includegraphics[width=\linewidth]{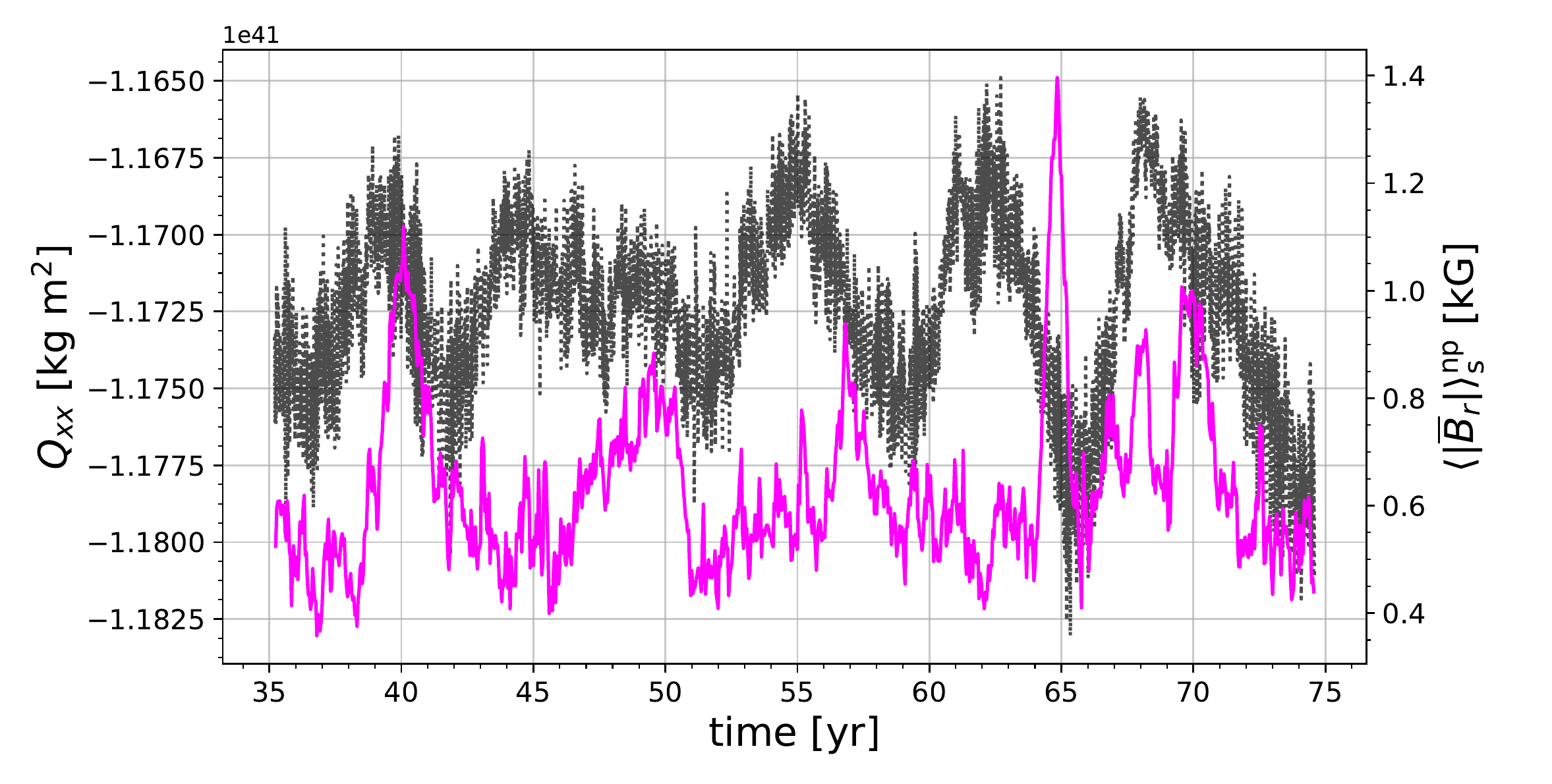}
\caption{Time evolution of the gravitational quadrupole moment
  component $Q_{xx}$ (black-dotted line) in \textit{run3x} together with the
  absolute value of the azimuthal-average of the radial magnetic field
  near the surface averaged over the northern hemisphere (magenta-solid
  line). The variations of $Q_{xx}$ can be interpreted as a reaction
  to the changes of the magnetic field intensity (see text). }
\label{fig:3-ns-q-bx}
\end{figure}

\begin{figure}
\includegraphics[width=\linewidth]{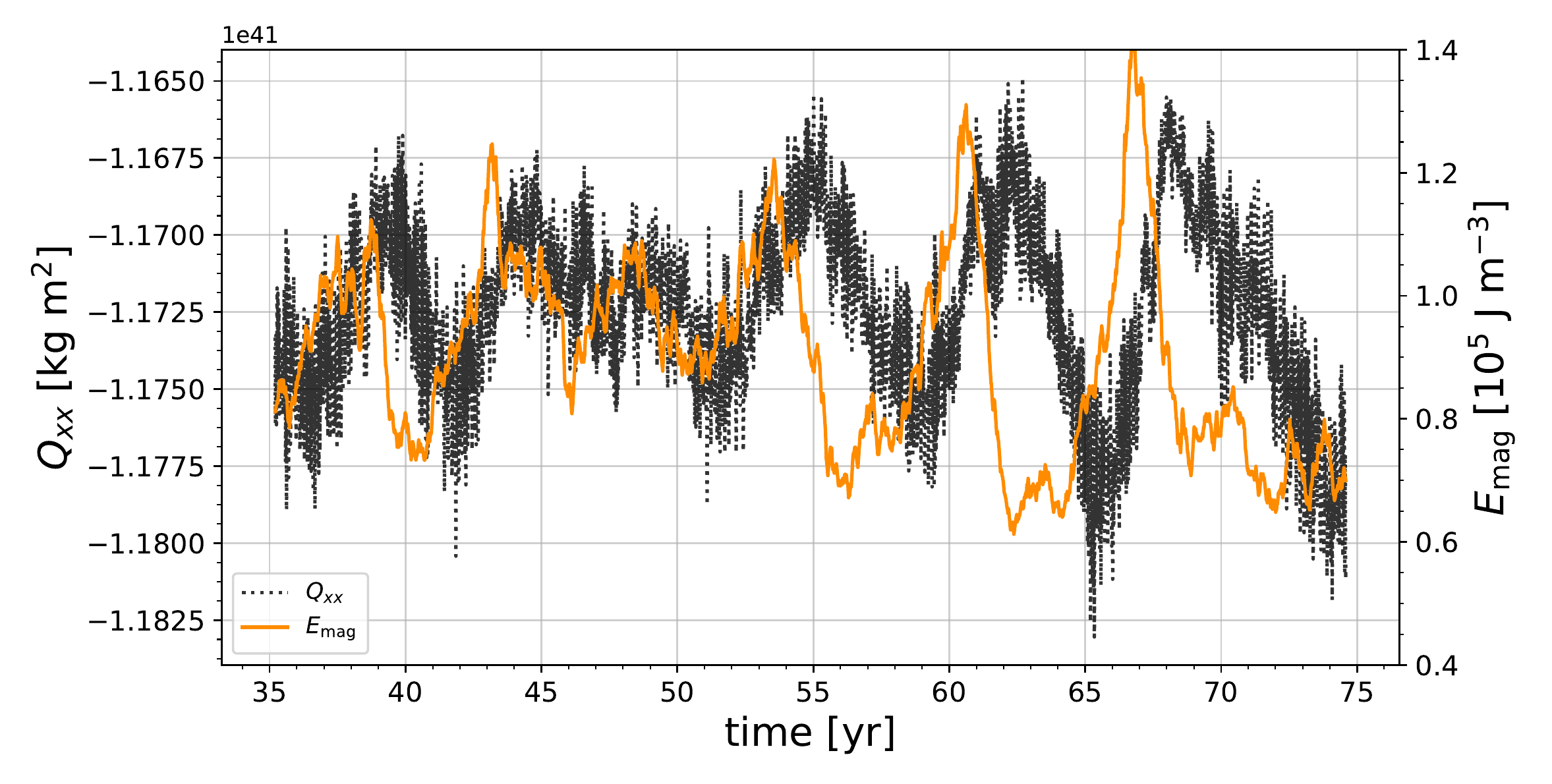}
\includegraphics[width=\linewidth]{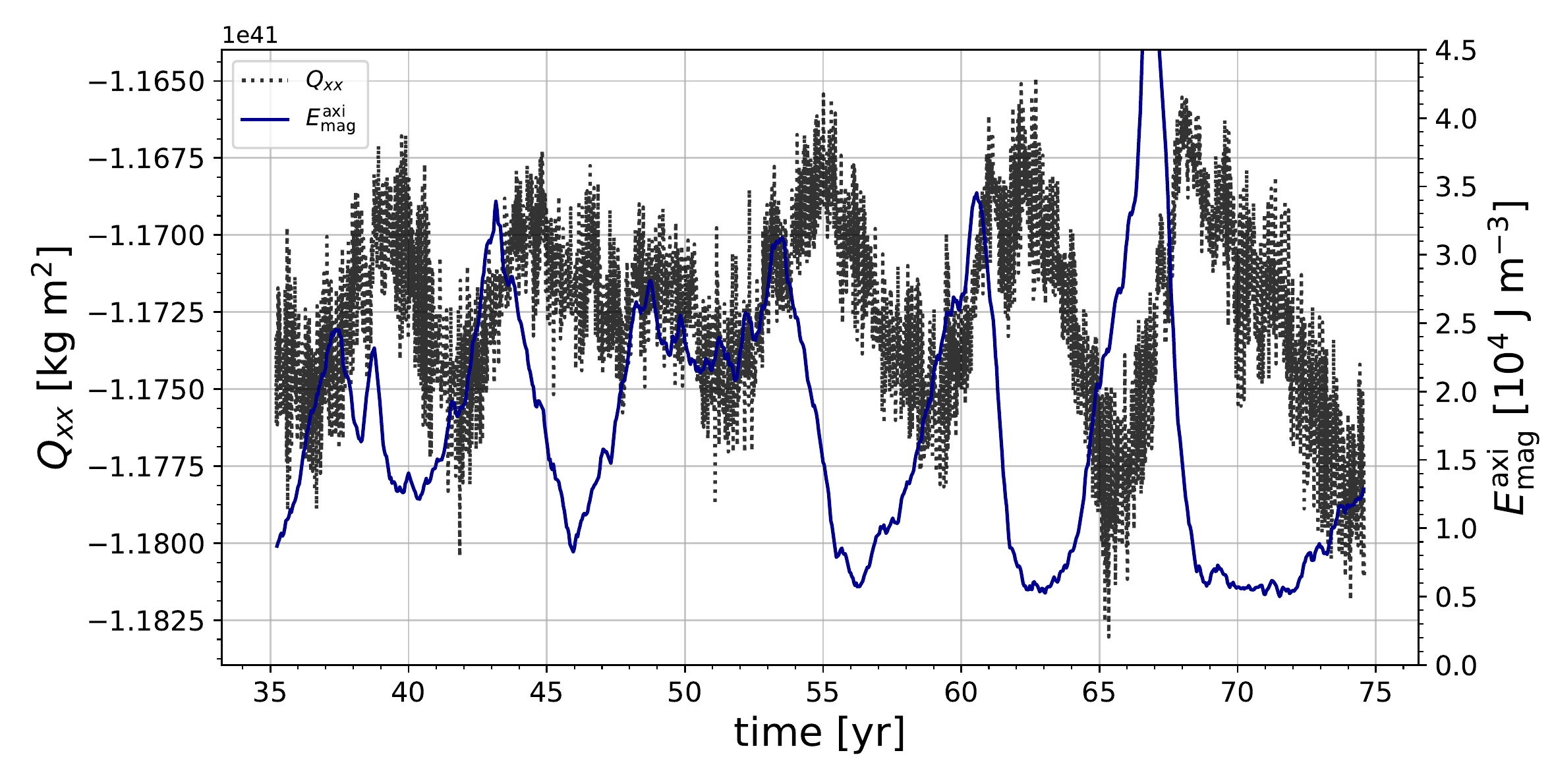}
\caption{Time evolution of the total magnetic energy (top panel) and the axisymmetric magnetic energy (lower panel), compared to the evolution of the gravitational quadrupole moment $Q_{xx}$ (black line) in run3x.}
\label{fig:pEmag_Qxx_run3x}
\end{figure}

We compare the evolution with the average radial magnetic field near the
surface averaged over the northern hemisphere in Figure
\ref{fig:3-ns-q-bx}. We can see peaks of the magnetic field and how
they relate to the quadrupole moment. The first peak of the magnetic
field at $t = 40$~years can be related to the minimum of $Q_{xx}$ at
$t = 41.7$ years. Then, the continuous increase in the magnetic field
intensity from $t = 45$~years to $t = 50$~years is reflected in a
decrease of $Q_{xx}$ starting at the 45~years mark to $t =
51$~years.  {This shows the direct impact of magnetic field to the
  overall density distribution of the star. We also compare the
  evolution of} $Q_{xx}$ {directly with the
  evolution of the total and axisymmetric magnetic energies in Figure~\ref{fig:pEmag_Qxx_run3x} top and bottom panels. Here we see a close anticorrelation between the magnetic energy and $Q_{xx}$ in both panels, and a time lag might also be present.}

\begin{figure}
\includegraphics[width=\linewidth]{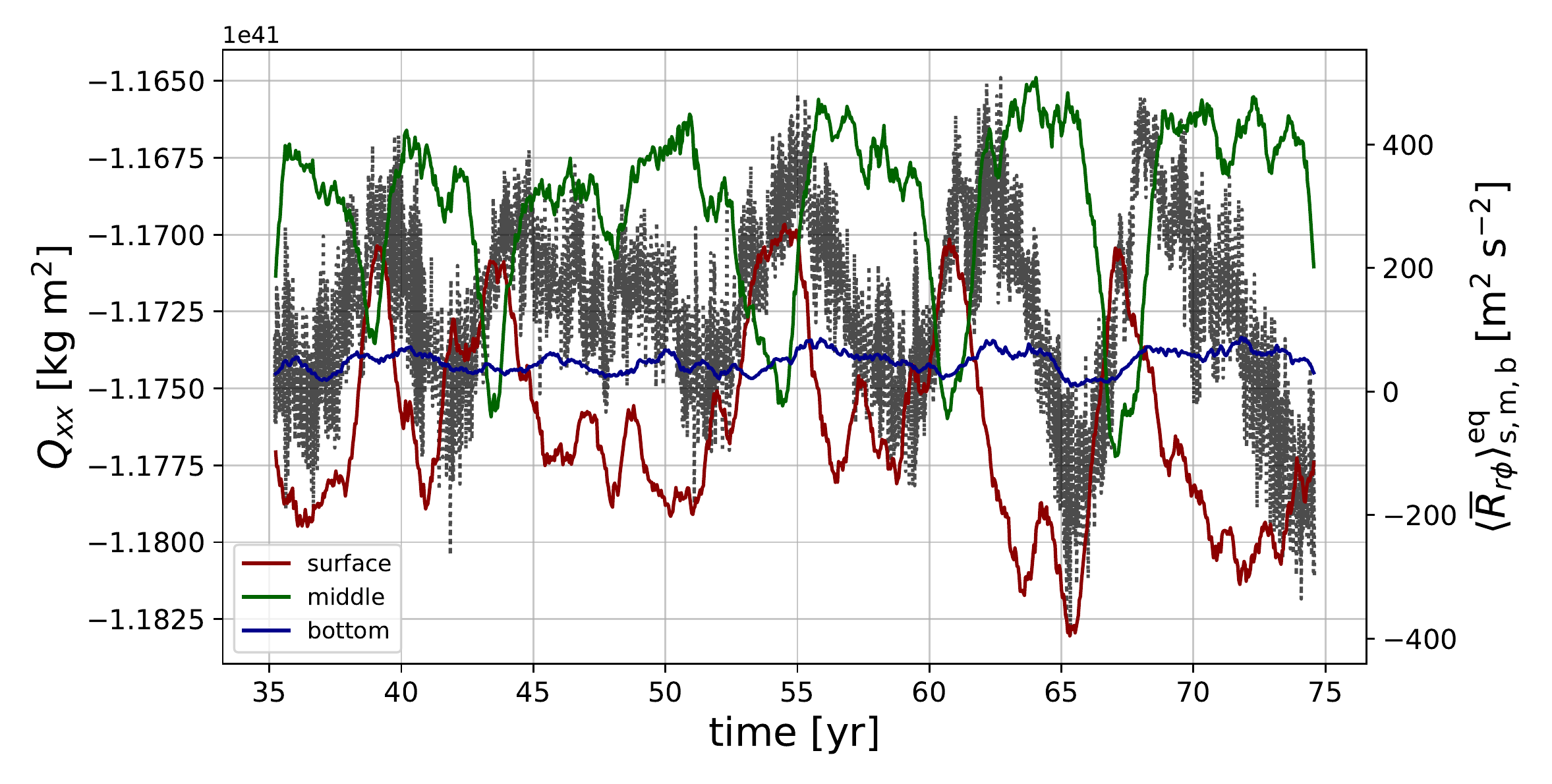}
\caption{Time evolution of the gravitational quadrupole moment
  component $Q_{xx}$ (black line) in \textit{run3x} together with the mean averaged of the
  Reynolds stress component $\overline{R}_{r\phi}$ at the equator in the surface
  (red), middle (green), and bottom (blue).}
\label{fig:3-eall-q-Rxz}
\end{figure}

Now, we explore the correlation between the Reynolds stress tensor
component $\overline{R}_{r\phi} = \overline{u^\prime_r
  u^\prime_\phi}$, where primes denote fluctuations from azimuthal
averages which are denoted by overbars, which is known to drive
differential
rotation \citep{R89,kapyla16}. This is shown in Figure
\ref{fig:3-eall-q-Rxz}. {The stress at the surface (middle) of the
  computational domain is correlated (anticorrelated) with the
  quadrupole moment. The stress at the bottom is weak and with a small
  contribution and weak correlation to $Q_{xx}$.}

\begin{figure}
\includegraphics[width=\linewidth]{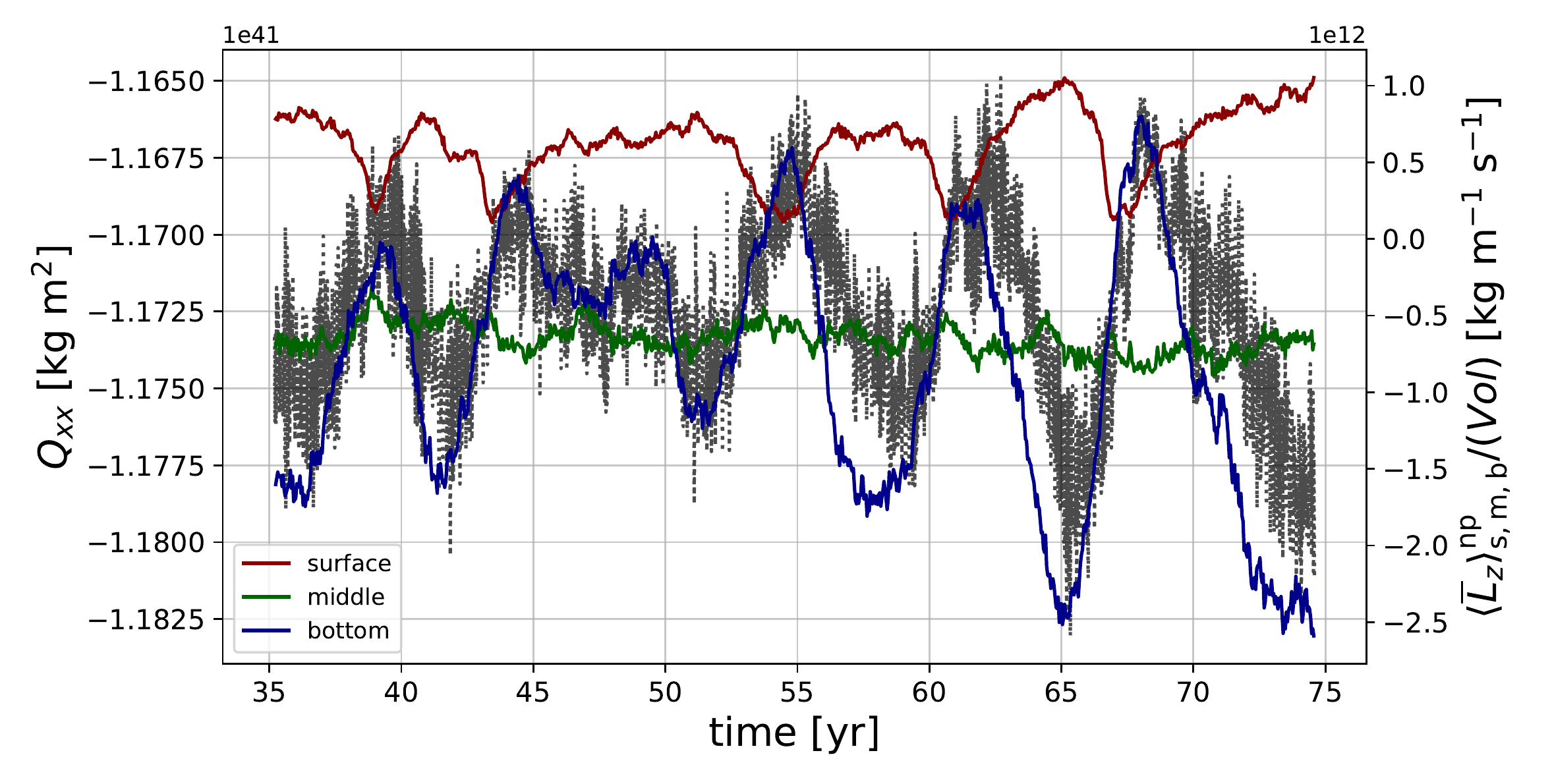}
\caption{Time evolution of the gravitational quadrupole moment
  component $Q_{xx}$ (black-dotted line) together with the angular
  momentum per unit volume averaged over the northern hemisphere at
  the surface (red), middle (green), and bottom (blue) for \textit{run3x}.}
\label{fig:3-nsmb-q-L-z}
\end{figure}

Finally, we study how the angular momentum {$L_z = \overline{\rho}
\varpi^2 \overline{\Omega}$, where $\varpi=r\sin\theta$} is related
to $Q_{xx}$. In
Figure~\ref{fig:3-nsmb-q-L-z}, we plot the angular momentum per unit
volume averaged over the northern hemisphere at the surface (red), middle
(green), and bottom (blue). While the angular momentum itself
will not directly affect the stellar structure through the centrifugal force,
it changes due to the Reynolds stress and shows a strong correlation
here with the change of the quadrupole moment.

From this figure we see that the outer layers carry less angular
momentum than the inner ones, and at the surface there is an anti-correlation between the absolute value of $Q_{xx}$ and the absolute value of $L_z$. The angular momentum at the surface further appears to be anti-correlated with the angular momentum in the middle and at the bottom, thus indicating an internal redistribution.

%
%
%
%
%
%

\subsubsection{Gravitational quadrupole moment evolution}\label{subsec:3-qxx}
 
{We return to Figure~\ref{fig:3-qall} to analyze the time
  evolution of quadrupole moment. We are interested in variations on
  timescales longer than the hydrodynamic oscillations with a period
  of $\sim$0.18 years, see Section~\ref{hydro}.}
The variations in $Q_{xx}$ are not strictly periodic. {For
  example, there is an episode between $t\sim44$~yr to $t\sim52$~yr
  where it takes clearly more time to reach a local
  maximum. Furthermore, $Q_{xx}$ reaches a global minimum at $t=65$~yr
  which is clearly lower than those that precede it}. This behaviour
is to be expected to a
certain degree as the full set of MHD equations is highly non-linear.
Overall, these fluctuations have a period of $\sim 5\ldots 6$
years and semi-amplitudes of $\mathbf{\sim1\times10^{39}}$ kg~m$^{2}$.

Bearing in mind the necessary rescaling to obtain astrophysical values
(see section \ref{rescaling}) and that we are modeling a solar mass
star, we can use the results from our simulations to estimate the
impact in V471~Tau. Following \citet{applegate92}, the variations in
the binary period are related to variations in the quadrupole moment
via
\begin{equation}
 \frac{\Delta P}{P} = -9\left(\frac{R}{a_\textnormal{bin}}\right)^2\frac{\Delta Q_{xx}}{MR^2},
\end{equation}
or,
\begin{equation}\label{eq:periodvar}
 \frac{\Delta P}{P} = -9\frac{\Delta Q_{xx}}{M\,a_\textnormal{bin}^2},
\end{equation}
where $M$ and $R$ are the mass and radius of the magnetically active star,
and $a_\textnormal{bin}$ is the binary separation. We take the
$Q_{xx}$ semi-amplitude as
\begin{equation}
 \Delta Q_{xx} = 1\times10^{39}\,\rm{kg~m}^{2}
\end{equation}
and adopt a binary separation of 3.3 $R_\odot$.  This result is consistent with the semi-analytic model by \citet{Volschow2018}, adopting fluctuations of about $10\%$ in the turbulence and magnetic field. We therefore obtain
\begin{equation}
 \frac{\Delta P}{P} = 8.4\times10^{-10}.
\end{equation}
Furthermore,
\begin{equation}
 \mbox{O--C} = \frac{\Delta P}{P} \frac{P_{\rm mod}}{2\pi},
\end{equation}
where $P_{\rm mod}$ is the modulation period of the $O-C$ diagram \citep[see][]{applegate92}. Combining this equation with Equation (\ref{eq:periodvar}) yields
\begin{equation}
 \mbox{O--C} = 0.025\,\rm{s}.
\end{equation}
\citet{marchioni18} presented the most updated analysis of the
eclipsing times of V471~Tau. The authors reported two period
variations, one with $O-C =$ 151~s and $P_{\rm mod} = 35$~years. The
other contribution has a semi-amplitude of $O-C =$~20~s and a
modulation period of $P_{\rm mod} = 9.7$~years. The semi-amplitude
obtained from the simulations in this case is thus much lower than
observed. However, we note that the rotation rate is different
than in V471~Tau, and also the stellar mass may not be exactly the
same. Indeed, more promising results will be obtained with the fast
rotator in the next section.
%
%
%
%
%
%

\subsection{The case of the fast rotator (\textit{run20x})}\label{fast}

{We now investigate the evolution of a simulation with twenty times
solar rotation. This case is characterised by ${\rm
  Ta}=2.55\times10^9$, ${\rm Co}=59.7$, ${\rm Re}={\rm Re}_M=21$,
${\rm Pr}_M=1$, and ${\rm Pr}_{\rm SGS}=2.5$. Also this simulation lies
within the parameter regime explored by \citet{viviani18}.}

\subsubsection{Overview of convective and magnetic states}\label{subsec:20x-surface}
\begin{figure}
\includegraphics[width=\linewidth]{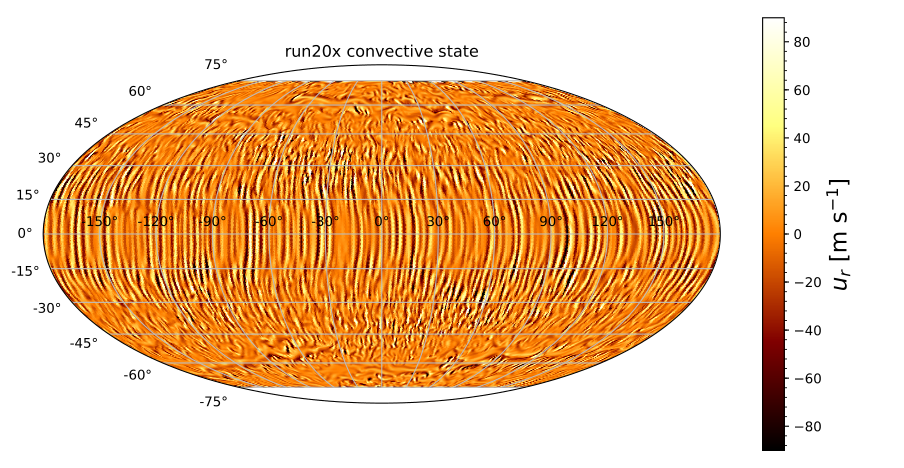}
  \caption{Mollweide projection of the radial velocity near the surface for \textit{run20x}. Colorbars are cut to improve visualization.}
\label{fig:20-state-convective}
\end{figure}

Figure \ref{fig:20-state-convective} shows the near-surface radial
velocity from \textit{run20x}. Also here, banana cells here are present at the
equator, but with a decreased azimuthal extent in comparison to
\textit{run3x}. At higher latitudes the size of the convection cells is also
reduced. This decreasing size of convection cells as the rotation
increases is in accordance with linear stability analysis
\citep[e.g.][]{Chandra61} {and earlier numerical simulations
\citep[e.g.][]{viviani18}.} The average convective velocity is
19.4~m~s$^{-1}$, with
extrema of 700~m~s$^{-1}$ and $-561$~m~s$^{-1}$.

\begin{figure}
\includegraphics[width=\linewidth]{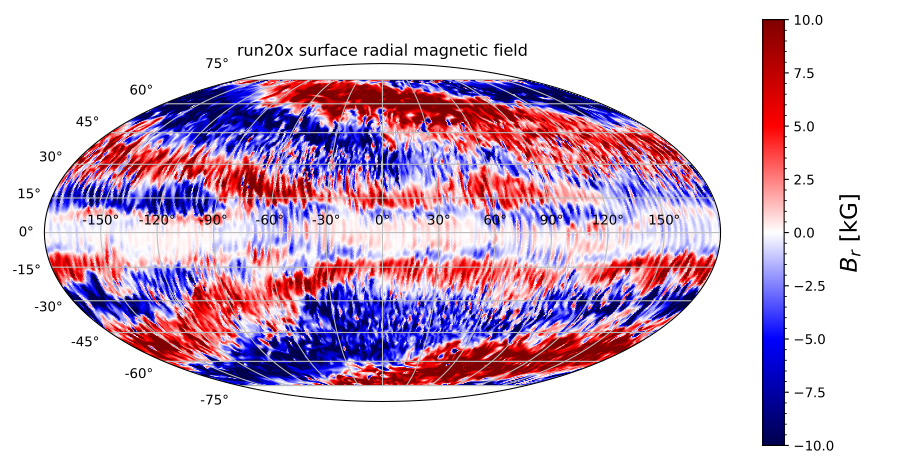}
\caption{Mollweide projection of the radial magnetic field near the surface for \textit{run20x}. Colorbars are cut to improve visualization.}
\label{fig:20-state-br}
\end{figure}

Figure~\ref{fig:20-state-br} shows the near-surface radial magnetic
field at the end of \textit{run20x}. The radial magnetic field differs from
that of \textit{run3x} in that it is stronger and more organized. The rms
radial magnetic field is 4.5 kG, i.e. 1.8 times stronger than in \textit{run3x}. The
extrema are about $\pm$~90~kG, as in \textit{run3x}. These large-scale
magnetic structures can cover the whole surface of the star.
A possible explanation is that convection in the rapidly rotating run
is less supercritical in terms of its Rayleigh number because the
values of $\nu$ and $\chi_{\rm SGS}$ remain the same as in \textit{run3x}. Thus
the flows and magnetic fields in \textit{run20x} are more laminar than in
\textit{run3x}.

\subsubsection{Overview of the magnetic field evolution}\label{subsec:20x-overview}

\begin{figure}
\includegraphics[width=\linewidth]{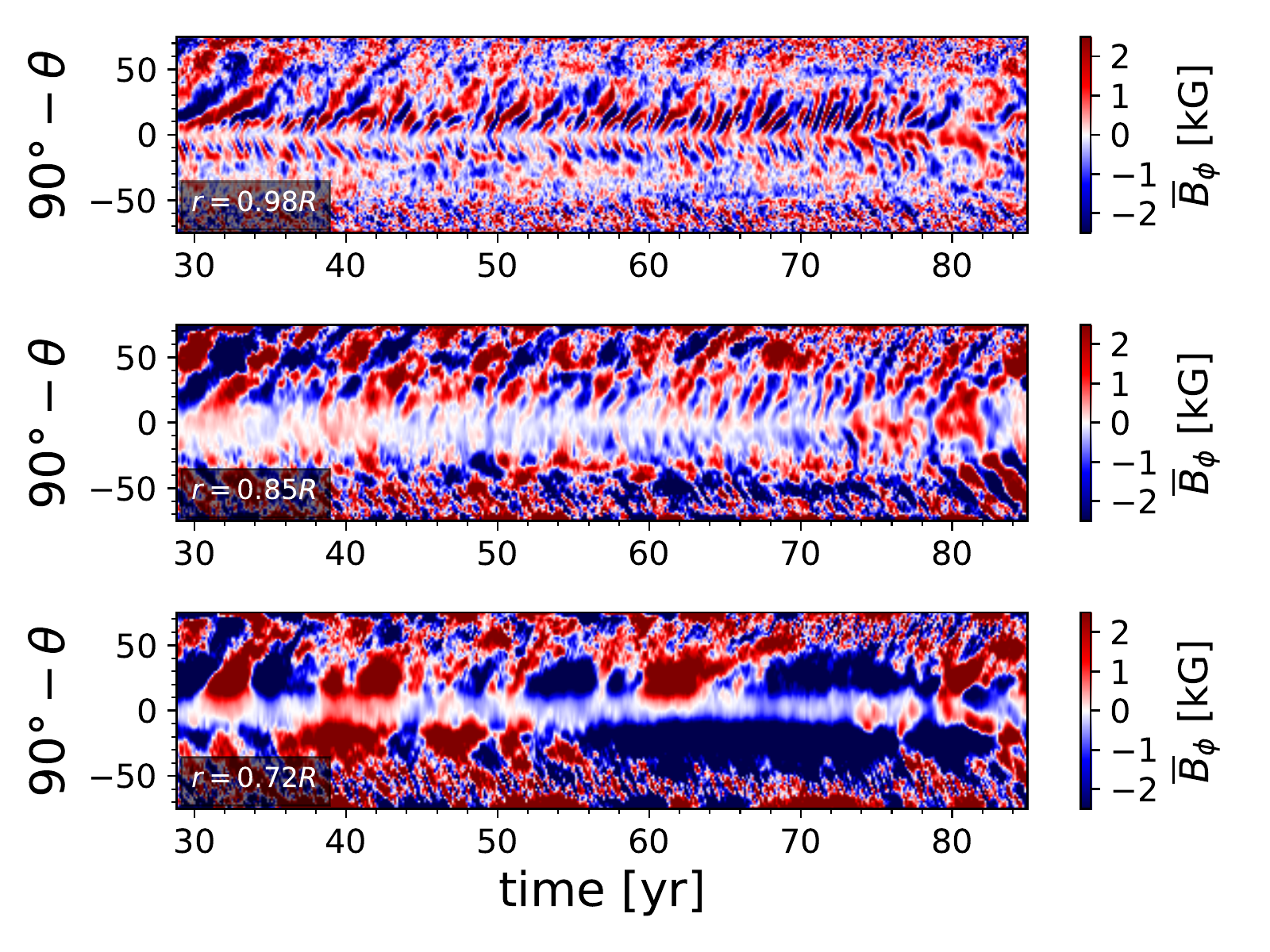}
\caption{Time evolution of the mean azimuthal toroidal magnetic field
  $\overline{B}_\phi$ for \textit{run20x}. A poleward migration of the magnetic
  field is clearly seen at the surface and middle of the domain. Near
  the equator there is a hemispheric wave operating on the
  northern hemisphere at latitudes between $\sim 5^\circ$ to $\sim
  50^\circ$. This hemispheric asymmetry is decreasing toward the end
  of the simulation. The
  colorbars are cut at $\pm$2.5 kG for better visualization.}
\label{fig:20-butterfly-bz}
\end{figure}

We follow here the same approach as in the case of the slow
rotator. Figure~\ref{fig:20-butterfly-bz} shows the {time
  evolution of the} mean toroidal
magnetic field, i.e. butterfly diagram, at three depths labeled at the
lower left corner of each panel. The mean magnetic field shows a more
complex behavior than in \textit{run3x}. At the bottom of the domain the dynamo
solution is
cyclic everywhere in the
beginning. The maximum amplitudes are $\pm$12 kG. At later times there
is a quasi-stationary axisymmetric magnetic field from 57 years to 76
years, covering most of the southern hemisphere. The dynamo solution
at the middle of the domain is
persistently cyclic with a poleward migration. Here the extrema of the
magnetic field are $\pm$8 kG. At the surface there is a poleward
dynamo wave near the equator with extrema of $\pm$5
kG. This poleward mode is clearly more coherent on the northern
hemisphere and can be seen at latitudes between $\sim 5^\circ$ to
$\sim 50^\circ$, whereas a higher frequency wave on the southern
hemisphere can be seen only very near the equator. The amplitude of
the axisymmetric magnetic field is also slowly decreasing during the
simulation. The absence of a
strong toroidal magnetic field near the surface is due to the radial
field boundary
condition \citep[see][]{kapyla16, warnecke16}.

\begin{figure}
\includegraphics[width=\linewidth]{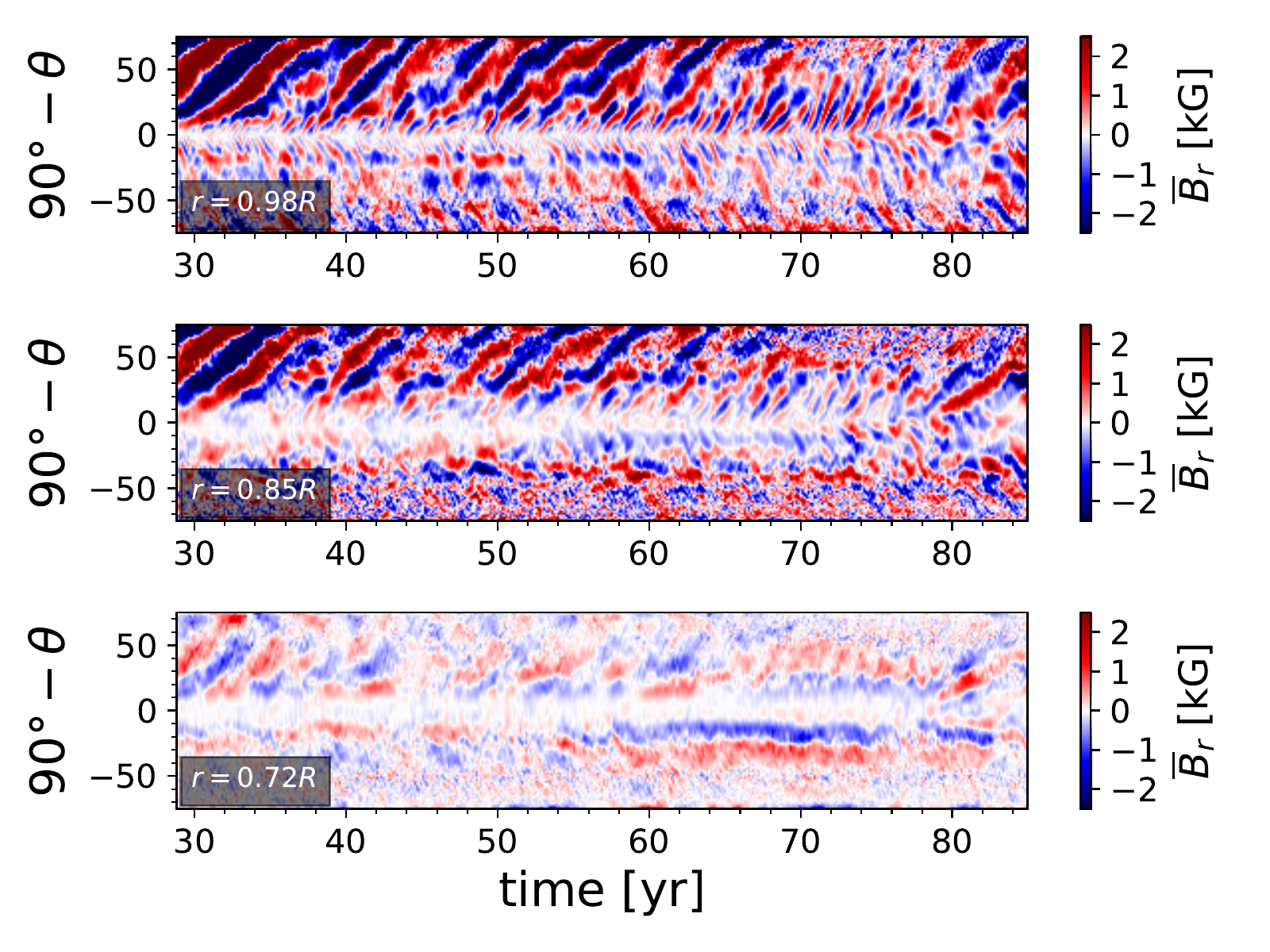}
\caption{Time evolution of the mean radial magnetic field
  $\overline{B}_r$ for \textit{run20x}. The hemispheric dynamo wave is
  clearly seen at the surface and middle of the domain. It is also
  seen that the magnitude of the axisymmetric field decreases until
  around 70~years.}
\label{fig:20-butterfly-bx}
\end{figure}

A time-latitude diagram for $\overline{B}_r$ is shown in Figure
\ref{fig:20-butterfly-bx}. Here, the presence of a hemispheric
dynamo wave with decreasing amplitude in time is clearly visible and
the magnetic fields have a poleward migration. At early times, the
extrema {at the surface (bottom) is $\pm$20~kG ($\pm$8~kG)}. The
hemispheric asymmetry disappears in the period between 68 to 80 years
and the extrema {near the top (bottom) is $\pm$4~kG ($\pm$3~kG),}
respectively.
The behaviour is quite different from the case of \textit{run3x}. This
is because the excited dynamo mode depends on the rotation rate of the
simulation \citep[see e.g.][]{viviani18}. The major differences in the
magnetic field evolution between \textit{run3x} and \textit{run20x} is that first, the
intensity of $\overline{B}_r$ in the latter is larger by a factor of 2
at the surface. Second, the overall intensity of the {azimuthally
  averaged} magnetic field in
the latter is decaying, whereas in \textit{run3x} it remains {
  roughly} constant. And third, the magnetic field is migrating
virtually
everywhere in \textit{run20x}, whereas the migrating component was found to be
subdominant in \textit{run3x} where a strong quasi-steady field is present at
all times (see Figure
\ref{fig:3-butterfly-bx}). We note in summary that the behaviour of
the magnetic field is considerably more complex in the rapidly
rotating case.

\subsubsection{Origin of the $Q_{xx}$ fluctuations}\label{subsec:20-q-origin}

\begin{figure}
\includegraphics[width=\linewidth]{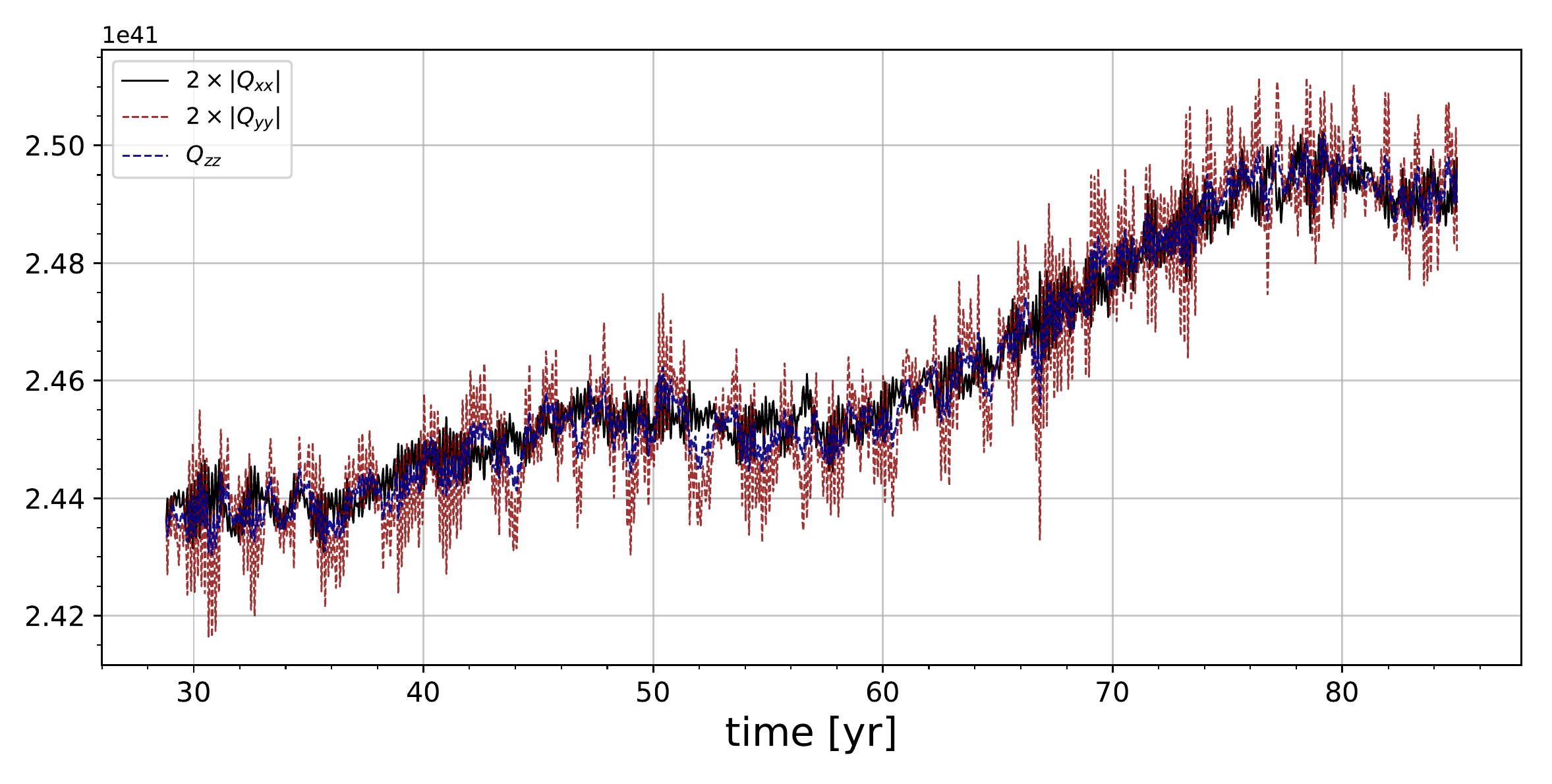}
\caption{Time evolution of the gravitational quadrupole moment components $2|Q_{xx}|$, $2|Q_{yy}|$ and $Q_{zz}$ in \textit{run20x}. Apart from differences of a factor of $\sim2$, the components follow the same overall trend, with short-term differences in the high frequency oscillations.}
\label{fig:20-qall}
\end{figure}

\begin{figure}
\includegraphics[width=\linewidth]{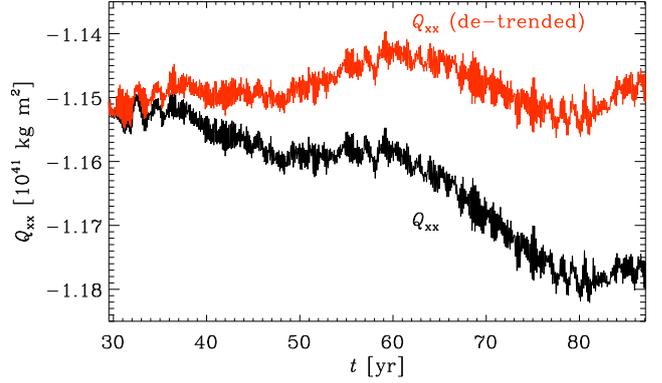}
\caption{Total and de-trended $Q_{xx}$ in black and orange, respectively, for \textit{run20x}.}
\label{fig:20-q-detrend}
\end{figure}

\begin{figure}
\includegraphics[width=\linewidth]{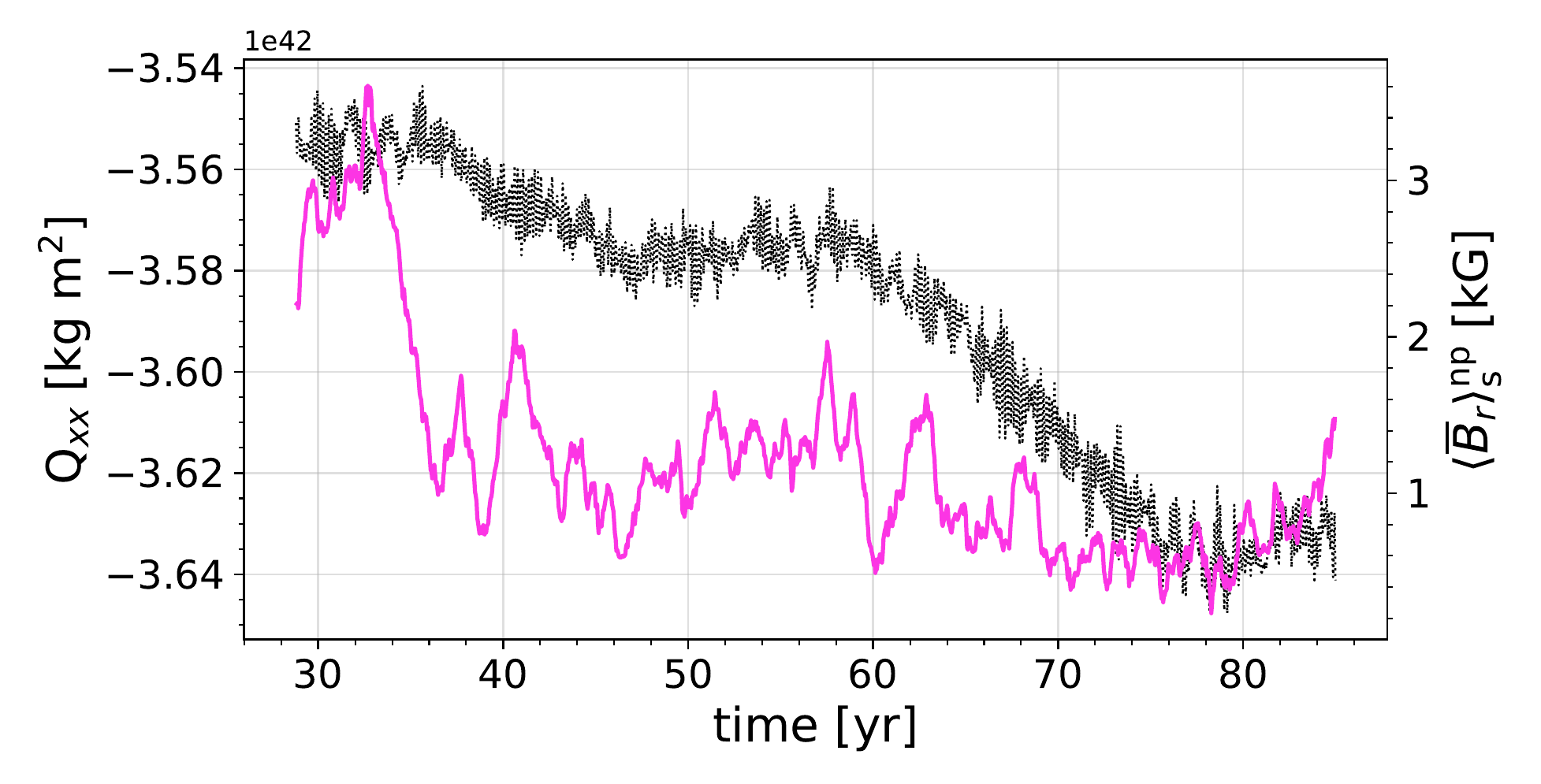}
\includegraphics[width=\linewidth]{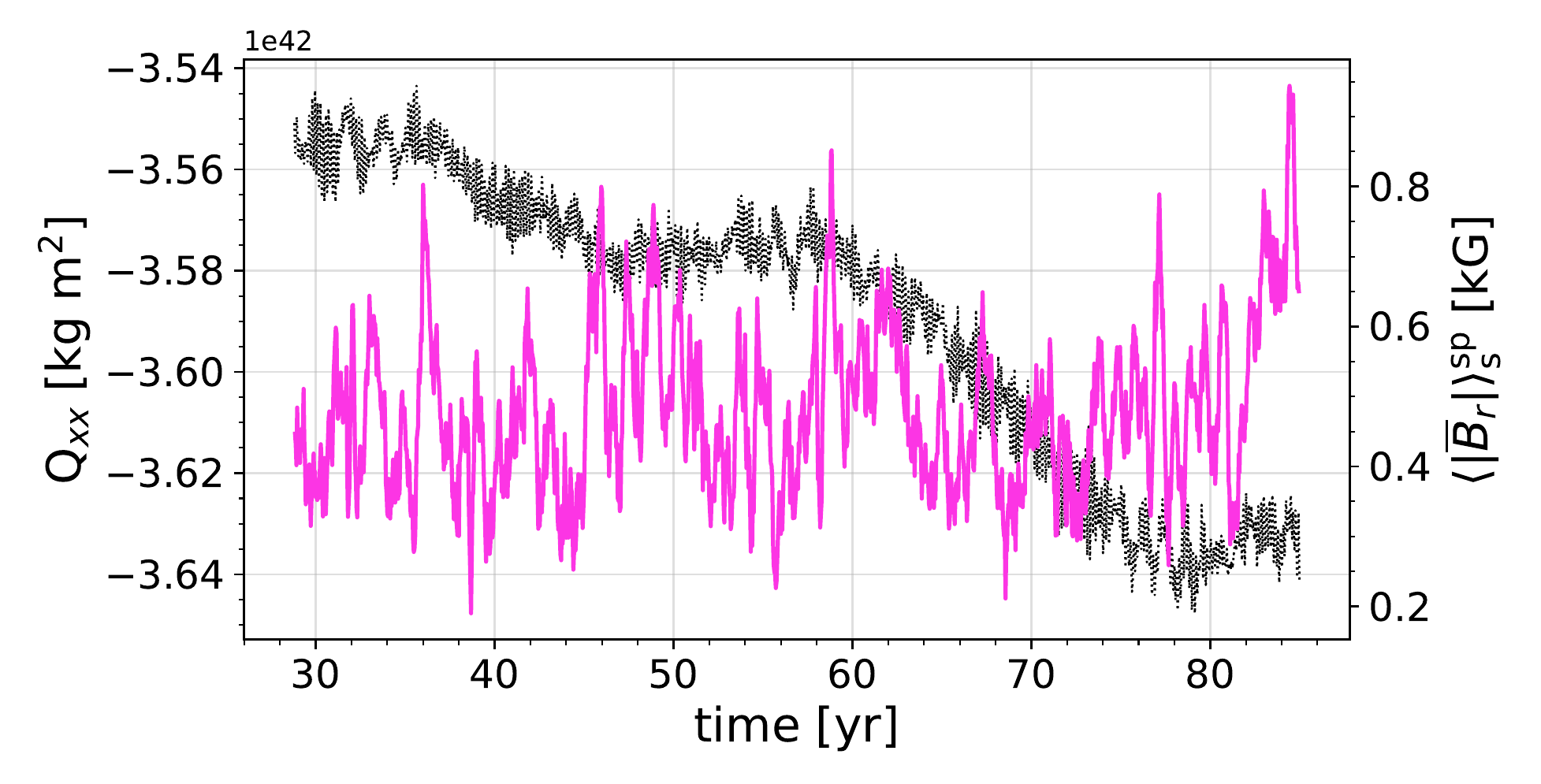}
\includegraphics[width=\linewidth]{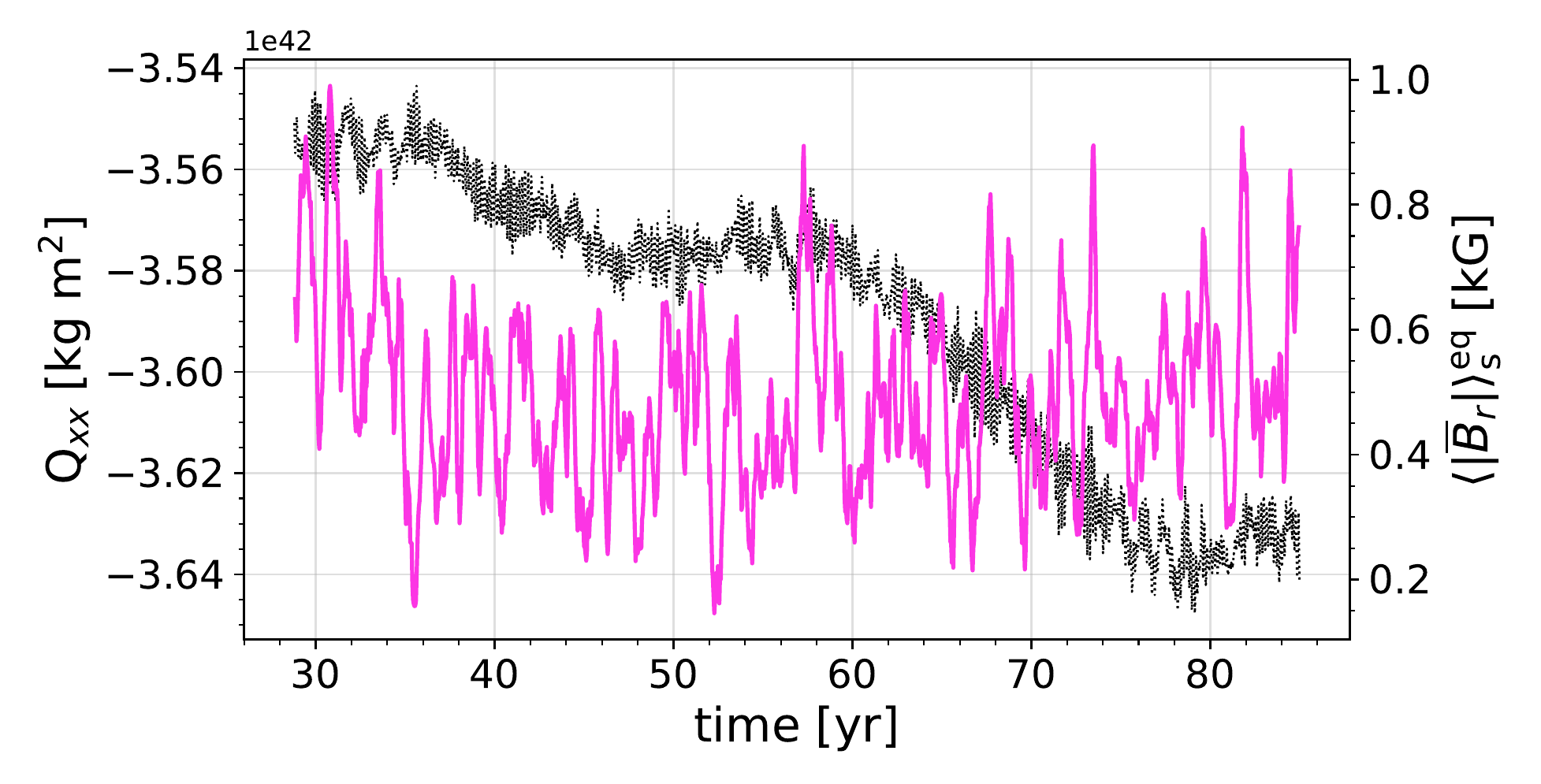}
\caption{Time evolution of the gravitational quadrupole moment component $Q_{xx}$ (black-dotted line) for \textit{run20x} together with the absolute value of the mean radial magnetic field  averaged at the north pole (top panel) in the surface of the domain (magenta-solid line). The mid panel shows the comparison with the mean radial magnetic field averaged at the south pole in the surface of the domain, in the bottom panel the average was taken at the equator in the surface of the domain.  }
\label{fig:20-ns-q-bx}
\end{figure}

Analogously to the slowly rotating case, we explore the origin of the
quadrupole moment
variations. {The time evolution of the diagonal elements $Q_{xx}$,
$Q_{yy}$, and $Q_{zz}$ are shown in Figure~\ref{fig:20-qall} but we
again study $Q_{xx}$ bearing in mind that the result would be similar
for the other components}. In $|Q_{xx}|$, we see a
quasi-periodic variation on a timescale of about 30 years superimposed
with a longer-term trend. {The latter, which decreases
continuously the quadrupole moment, might be related to an incomplete thermal saturation of the stellar interior. For this reason we have de-trended $Q_{xx}$
by taking the difference between the endpoints of its time series and
substracting this linear trend from the data.
The resulting time series is shown in Fig.~\ref{fig:20-q-detrend}.}
We first compare {the total, non-de-trended} $Q_{xx}$ to the evolution of
the mean radial magnetic field averaged at the northern hemisphere at the
surface of the domain, depicted in Figure \ref{fig:20-ns-q-bx}. The
sharp decrease of $\langle\overline{B}_r\rangle^{\rm{np}}_{\rm{s}}$
reflects the change in the dominant dynamo mode (same as in
Figure~\ref{fig:20-butterfly-bx}). {The effect of this decrease on
  $Q_{xx}$ can also be clearly seen}.

\begin{figure}
\includegraphics[width=\linewidth]{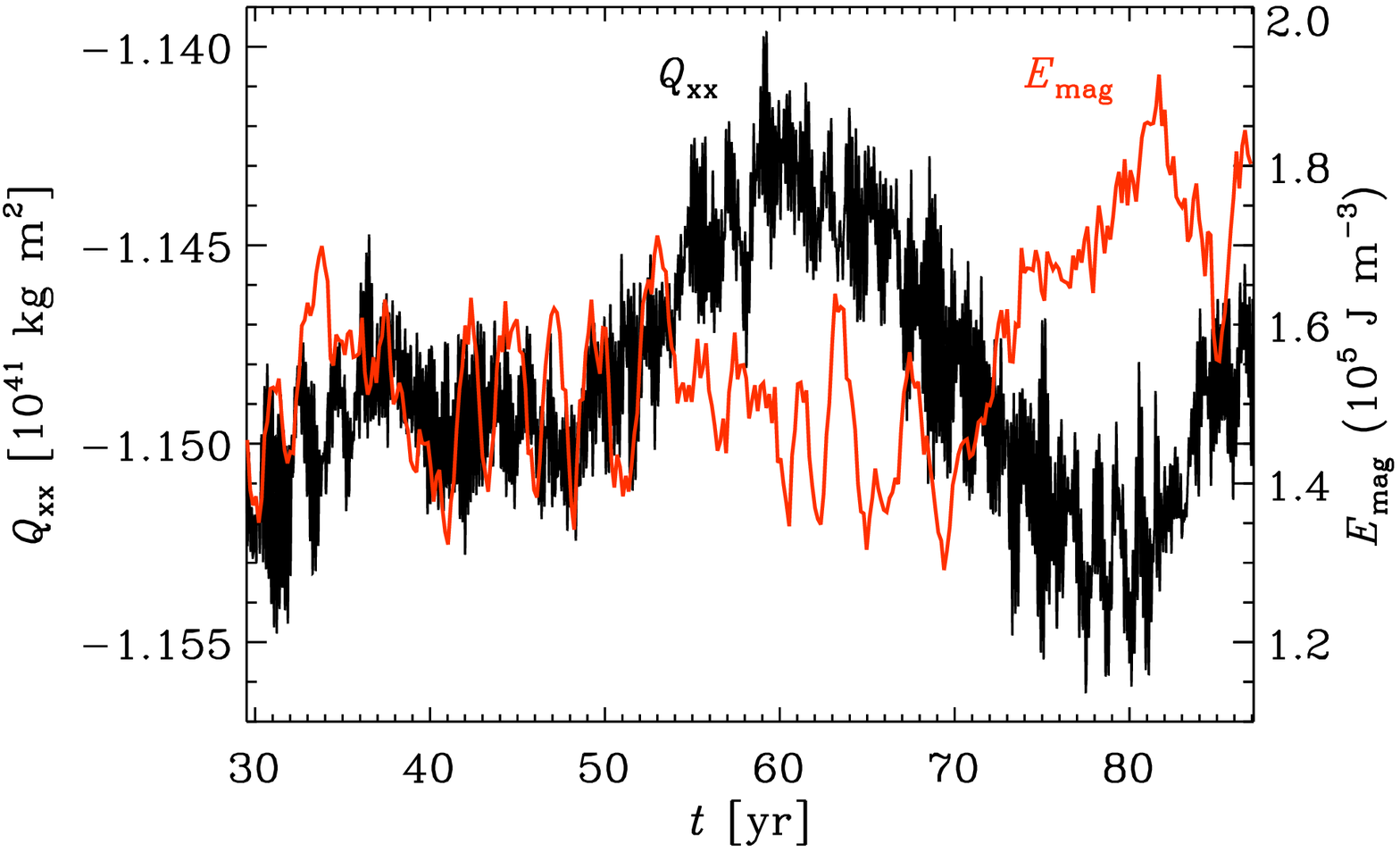}
\includegraphics[width=\linewidth]{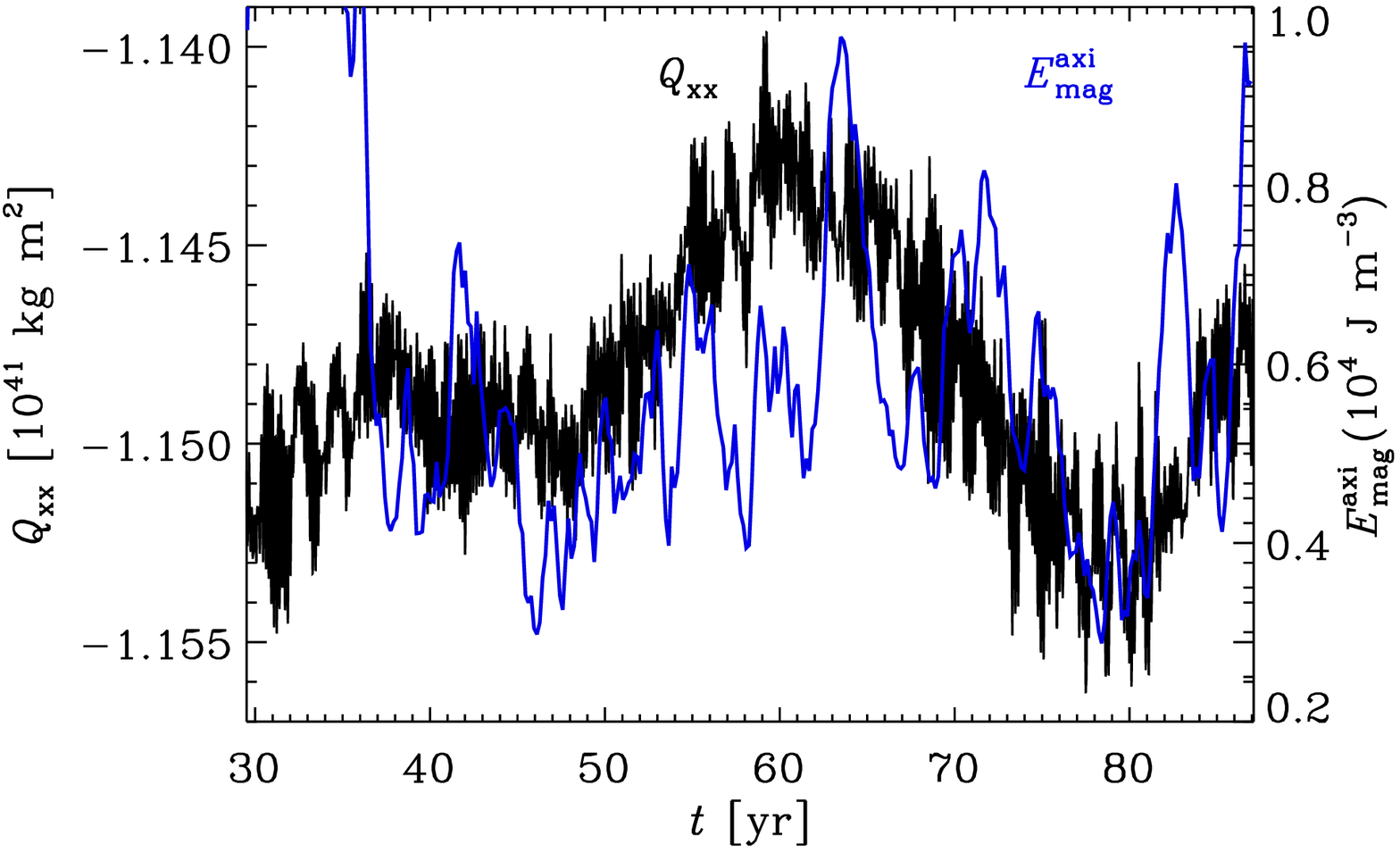}
\caption{Time evolution of the total magnetic energy {(top panel)} and the {axisymmetric} magnetic energy {(lower panel)}, compared to the evolution of the {de-trended} gravitational quadrupole moment $Q_{xx}$ (black line) in \textit{run20x}.}
\label{fig:20-Emag}
\end{figure}

This correlation is not seen when the same quantities are compared on
the southern hemisphere (see middle panel of
Figure~\ref{fig:20-ns-q-bx}). {It can also be
seen that the mean magnetic field is weaker by almost an order of
magnitude in the southern hemisphere around $t=32$~yr. The mean field
strengths evolve gradually such that they are equal around $t=70$~yr.
The mean fields at both hemispheres start to grow around the 80~year
mark}; see top panel of Figure
\ref{fig:20-ns-q-bx}) which coincides with the weak increase in
$Q_{xx}$. Finally, it can be seen from the bottom panel of Figure \ref{fig:20-ns-q-bx}
that the average magnetic field at the equatorial portion of the
surface of the star does not have significant variations nor
correlation with $Q_{xx}$. {We further compare the evolution of the \textit{de-trended} $Q_{xx}$
with the total magnetic and the azimuthal
magnetic energy in Figure~\ref{fig:20-Emag}. Here, especially for the azimuthal magnetic energy, the anticorrelation is
less pronounced than we previously found in run \textit{run3x}.} In the total magnetic energy, several maxima or minima
{show counterparts in the evolution of $Q_{xx}$ but only toward
  the end of the simulation, i.e.~$t\gtrsim 65$~yr}. Further, while
$Q_{xx}$ on
average is decreasing during the period investigated here, the total magnetic energy shows an average increase. The
azimuthal magnetic energy, on the other hand, shows a strong peak towards the beginning of the simulated period, and
subsequently remains {at a lower level with quasi-periodic
  fluctuations on a timescale of roughly 5 years in superposition with
  a possible longer term modulation.}

\begin{figure}
\includegraphics[width=\linewidth]{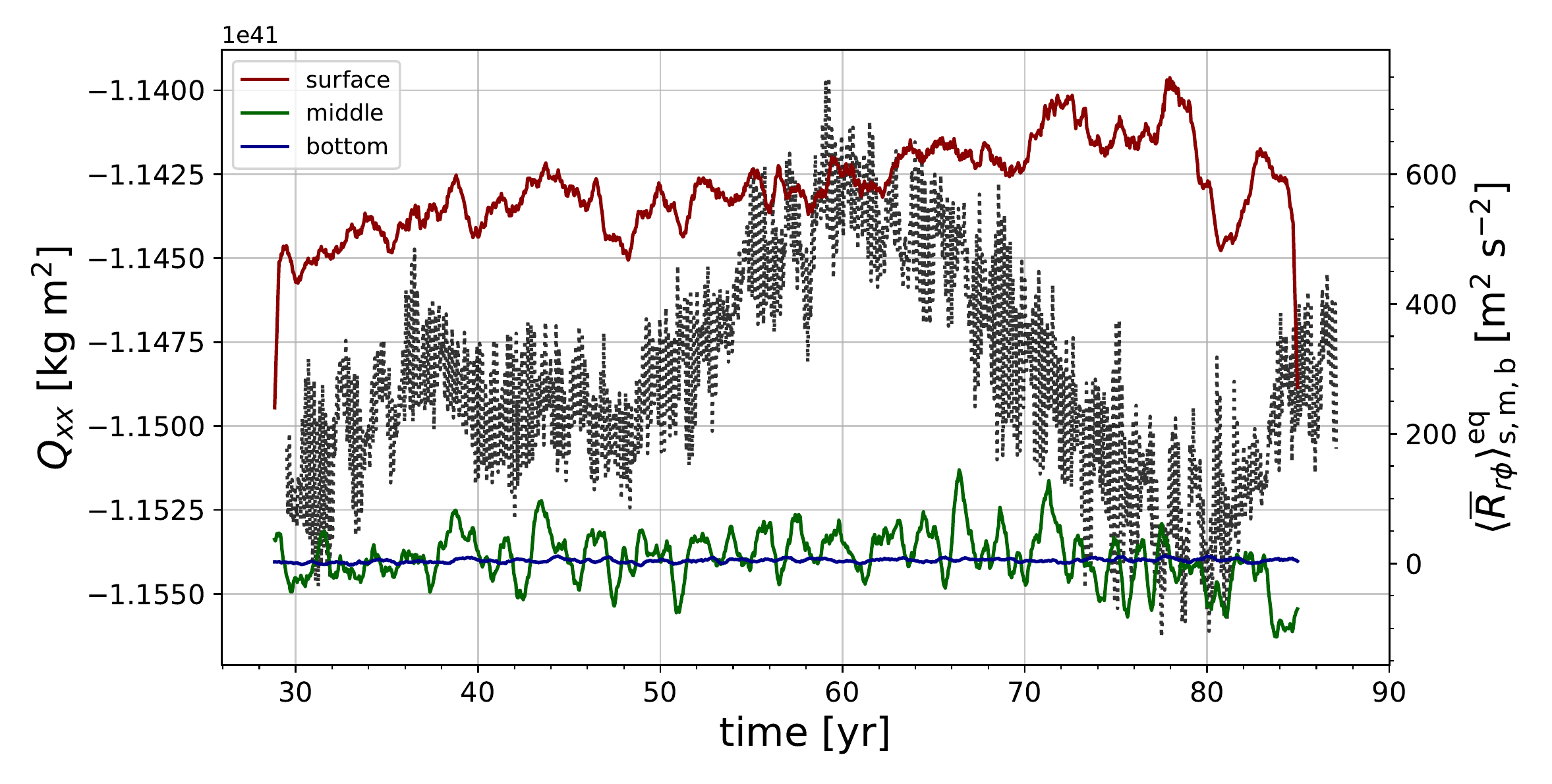}
\caption{Time evolution of the gravitational quadrupole moment
  component $Q_{xx}$ (black-dotted line) for \textit{run20x} together with the
  mean averaged Reynolds stress component $R_{r\phi}$ at the equator
  in the surface (red line), middle (green), and bottom
  (blue) of the domain.}
\label{fig:20-eall-q-Rxz}
\end{figure}

\begin{figure}
\includegraphics[width=\linewidth]{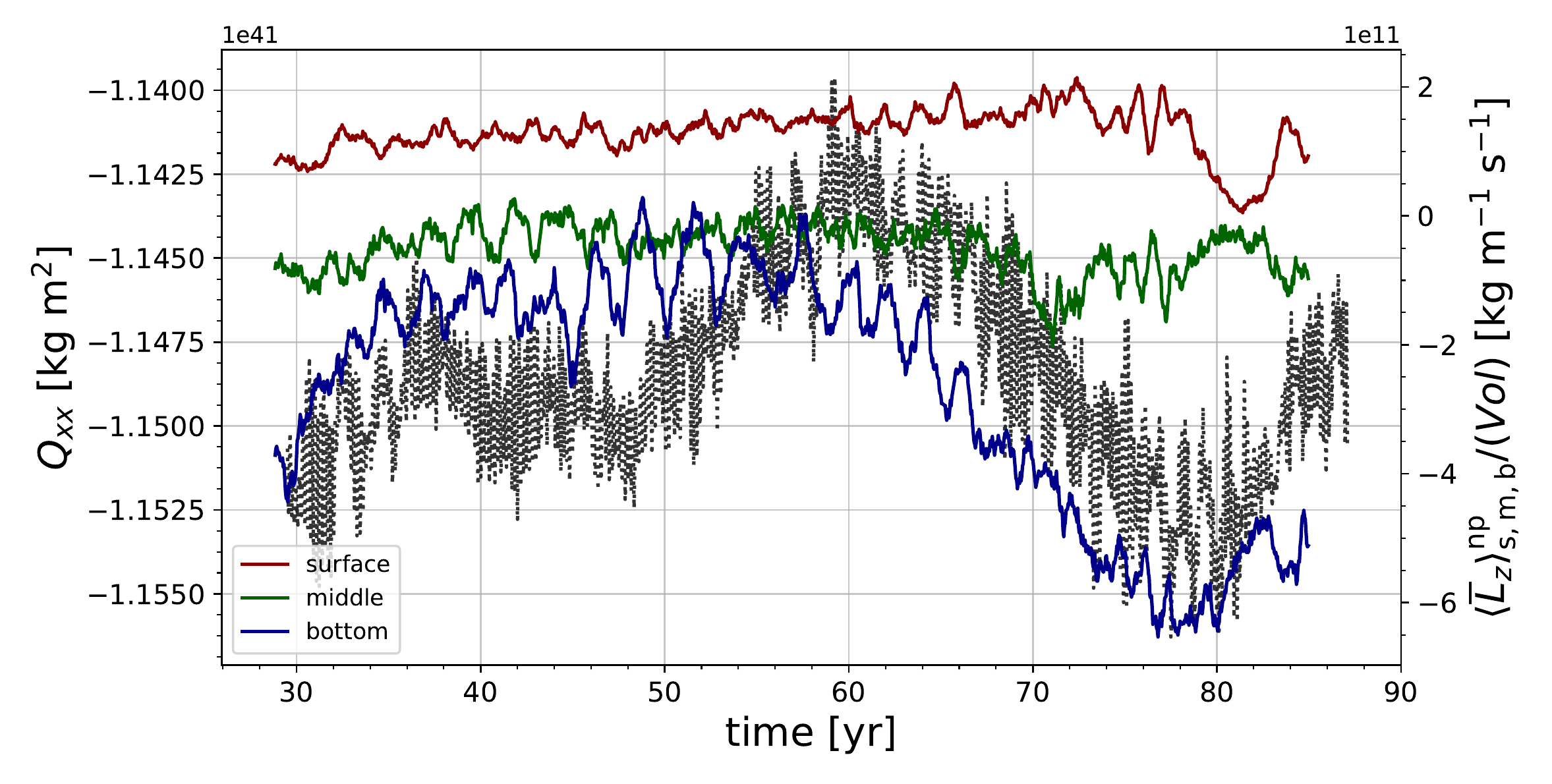}
\caption{Time evolution of the gravitational quadrupole moment
  component $Q_{xx}$ (black-dotted line) for \textit{run20x} together with the
  angular momentum per unit volume averaged over the northern hemisphere at the
  surface (red), middle (green), and bottom (blue) for the
  fast rotator. }
\label{fig:20-nsmb-q-L-z}
\end{figure}

As for \textit{run3x}, {we plot the $r\phi$ component of the
  Reynolds stress at the equator and at the three depths in
  Figure~\ref{fig:20-eall-q-Rxz}}. The average of
$\overline{R}_{r\phi}$ at the surface is steadily increasing while $Q_{xx}$ decreases. Meanwhile, the stress
at the deeper layers is approximately constant. {There is
  anti-correlation between $Q_{xx}$ and the stress at the equator near
  the surface in the latter part ($t\gtrsim 60$~yr) of the
  simulation}. We also show the
correlation between the quadrupole moment and the angular momentum at three depths in 
Figure~\ref{fig:20-nsmb-q-L-z}, though we note as before that this correlation is due to the effect of 
angular momentum redistribution by the Reynolds tensor, and unrelated to the centrifugal acceleration. 
While at early times $L_z$ at the bottom increases and $Q_{xx}$ remains constant, the correlation in 
both quantities after the 40~years mark is high.
{To recapitulate, we find that the evolution of the fast rotator
  is much more complex in comparison to the more slowly rotating
  counterpart and clear correlations between magnetic activity and
  quadrupole moment variations are visible only toward the end of the
  simulation. A significantly longer time series would be needed to
  quantify this effect more precisely. This is out of the scope of the
  current study and will be pursued elsewhere.}

\subsubsection{Gravitational quadrupole moment evolution}\label{subsec:20x-q}

The {purely hydrodynamic} oscillations in the quadrupole moments, particularly
$Q_{xx}$ (see also Sect. \ref{hydro}) are present, similarly as in
\textit{run3x}. The overall behaviour of the quadrupole moment in this run is
remarkably different from the case of \textit{run3x}, showing a more complex
behaviour. At the beginning from {the 29 to 37 year marks} $Q_{xx}$ remains
constant on average, apart from the presence of hydrodynamic oscillations with a
period of $\sim$0.18 years. After the 37~years mark, $Q_{xx}$ decreases gradually
from {
${-1.15\times10^{41}}$~kg~m$^2$ to 
$-1.16\times
10^{41}$~kg~m$^2$. After this, the behavior described above starts
again but now the decrease is stronger and starts at
60~years. $Q_{xx}$ changes from $-1.17\times10^{41}$~kg~m$^2$ to
$-1.18\times10^{41}$~kg~m$^2$. It is thus possible that we see here a
quasi-periodic oscillation superimposed with a longer term
trend. Analogously to the case of \textit{run3x} where we rescaled the
gravitational quadrupole moment (see section \ref{rescaling}), we take
the system parameters of the magnetically active component in the PCEB
V471~Tau and bear in mind that this run has a rotation rate and stellar
parameters similar to the magnetically active star in this system, but
now we take the maximum and minimum of $Q_{xx}$ to obtain
\begin{equation}
 \Delta Q_{xx} = 1.7\times10^{39}\,\rm{kg\,m}^2
\end{equation}
and adopt a binary separation of 3.3~$R_\odot$. Inserting this into Equation (\ref{eq:periodvar}) yields
\begin{equation}
 \frac{\Delta P}{P} = 1.40\times10^{-9}.
\end{equation}
Furthermore,
\begin{equation}
 O-C = \frac{\Delta P}{P}\frac{P_{\rm{mod}}}{2\pi}
\end{equation}
where $P_{\rm{mod}}$ is the modulation period of the O-C diagram semi-amplitude \citep[see][]{applegate92} with (\ref{eq:periodvar}). In our case $P_{\rm{mod}} = 50$ years. Thus
\begin{equation}
 O-C = 0.13\,\rm{s}.
\end{equation}
The semi-amplitude obtained from the simulations is {still lower than the observed value found by \citet{marchioni18}. Nevertheless, we also found that it has increased considerably compared to the slow rotator, by a factor of 5.2, while the rotation velocity has changed by roughly a factor of 6.7. As we are still a factor of 2.5 below the rotation velocity of V471~Tau, it is conceivable that another significant increase could be expected for the parameters of that system. In addition, we note that the centrifugal force is neglected in our simulations, which can be another relevant contribution.}

\section{Discussion and conclusions}\label{conclusions}

In this paper, we have studied the stellar quadrupole moment variations arising from magnetic activity through directly solving the 3D compressible non-ideal MHD equations with the {\sc Pencil  Code}.
We have run two {simulations of solar mass stars},
one with three times the solar rotation rate, and {the other} with
20 times solar rotation. {This is motivated by the fact that}
typical rotation rates in PCEBs are
considerably higher than for isolated stars. As a reference system, we
here consider V471~Tau, which has a {roughly solar mass}
secondary.

In the two simulations we have run, we see two very different
behaviours in the evolution of the magnetic fields and the quadrupole
moment. For the slow rotator, quasi-periodic oscillations in the
quadrupole moment, the magnetic field, the Reynolds stress and other
quantities can be distinguished easily. Meanwhile, for the fast
rotator the evolution is much more complex, which can also be seen in
the magnetic field evolution. The slow rotator has a relatively simple
magnetic field behaviour, showing a superposition of a strong
quasi-steady and a weaker migrating dynamo modes, whereas the fast
rotator has a significantly more
complex magnetic field evolution. It has a poleward migrating magnetic
field near the equator and a superposed hemispheric dynamo wave
operating only on the northern hemisphere. The latter is also
decreasing
its amplitude. {While the run has been evolved for a total of 90 years, it may not yet be in complete thermal saturation, which can give rise to the long-term trends that we observed. We therefore have de-trended the simulations to correct for such an influence, yielding then a clear anti-correlation with magnetic energy.}

We have established a link between the magnetic activity and the
gravitational quadrupole moment by means of the Reynolds stress
tensor, which will be affected by the magnetic dynamo due to its local
effect on the convective velocities. {There is an anticorrelation
between both the total and axisymmetric magnetic energies and $Q_{xx}$,
but we do not discard a time lag of the anticorrelation}.
While in the case of the slow rotator
it is relatively easy to observe, in the fast rotator case the
behaviour is much more complex, as it shows signs of a quasi-periodic
change, on which a global trend appears to be superimposed both for
the magnetic field and the quadrupole moment. The time line in our
simulations ($\sim55$~years) is larger than the observed timeline in
V471~Tau, while the observed timeline corresponds to about
$35$~years. {The expected O-C variation has increased considerably going from the slow to the fast
rotator, where a change by half an order of magnitude in the rotation velocity corresponds to a change by
a factor of 5.2 in the expected value of O-C. As even the fast rotator is a factor of 2.5 below the
rotation velocity of V471~Tau, another significant increase may be expected for the rotation velocity of
that system. We also note that the effect of the centrifugal force has been neglected so far, but it may
further enhance the O-C variations. The current simulations also
assume a fixed spherically symmetric gravitational potential. This
modeling choice is possibly also limiting the quadrupole moment
variations.}

Overall, we arrive at the following preliminary conclusions:
\begin{enumerate}
 \item the complexity of the evolution of $Q_{xx}$ is linked to the dynamo mode, angular momentum evolution, and Reynolds stress tensor,
 \item {$Q_{xx}$ is anticorrelated to the total and axisymmetric magnetic energies,}
 \item the numbers of the $O-C$ amplitude and $\Delta P/P$ depend on the overall magnetic field evolution and complexity,
 \item the angular momentum at the bottom of the convection zone is more correlated to $Q_{xx}$ than that near the surface,
 \item $\Delta Q_{xx}$ has a  {dependence} on stellar rotation.
\end{enumerate}

In spite of relevant uncertainties to be explored, we present here the
first analysis showing how the stellar quadrupole moment changes as a
function of time in compressible non-ideal MHD simulations. {We
  find}
strong evidence that magnetic effects can indeed produce such
variations, while pure hydrodynamical runs as presented in
section~\ref{hydro} {produce only} short-term variations on the
sound-crossing timescale. We believe that such simulations will be
important in the future to more quantitatively explore the effects of
magnetic activity in close binary systems, and to allow a better
understanding of the observed phenomena.

The variations in $Q_{xx}$ found here should be taken as indicative
rather than precise, as with the current computational power it is
impossible to approach the \textit{real} dimensionless parameters that
govern stellar plasmas. For example, the magnetic Prandtl number is
$1$ in the simulations whereas in the Sun it is $\sim 10^{-5}$. The
normalized flux in the bottom of the Sun is $\sim10^{-11}$ whereas in
the simulation it is highly enhanced with a value of
$3.2\times10^{-5}$. In the case of the Reynolds number this is more
severe, as in the Sun it ranges from $10^{12}$ to $10^{13}$ and in the
simulations we have ${\rm Re} \sim 21-71$. However, the simulations in
previous studies have proven to be successful in
reproducing some of the solar phenomena \citep[see e.g.][]{viviani18,
  kapyla16, kapyla13, kapyla12}. Further development of 3D MHD
simulations of fully-convective stars will prove to be of great
importance as we expect the Applegate mechanism to be an important
tool for studying M~dwarfs dynamos through eclipsing time variations.

To draw stronger conclusions, more simulations are required to explore the parameter space. In particular, exploring how $Q_{xx}$ depends on stellar rotation and mass is important as the magnetically active companion in PCEBs is rotating at a high fraction of their critical stellar rotation, which scales with the energetical feasibility of the Applegate mechanism \citep{Navarrete2018}. Also, fully-convective stars are expected to produce a higher amplitude of ${\Delta P/P}$ based on the models of \citet{Volschow2018}. Based on such simulations, eclipsing time observations may become a promising tool to probe stellar dynamos in the future.

\section*{Acknowledgements}
FHN acknowledges financial support from CONICTY (project code CONICYT-PFCHA/Magister Nacional/22181506). DRGS and FHN thank for funding through Fondecyt regular (project code 1161247) and through the ``Concurso Proyectos Internacionales de Investigaci\'on, Convocatoria 2015'' (project code PII20150171). R.E.M. and D.R.G.S. acknowledge FONDECYT regular 1190621 and the BASAL Centro de Astrof\'isica y Tecnolog\'ias Afines (CATA) PFB-06/2007. We acknowledge the Kultrun Astronomy Hybrid Cluster (projects Conicyt Programa de Astronomia Fondo Quimal QUIMAL170001, Conicyt PIA ACT172033, Fondecyt Iniciacion 11170268 and BASAL Centro de Astrof\'isica y Tecnolog\'ias Afines (CATA) PFB-06/2007) for providing HPC resources that have contributed to the research results reported in this paper. \{Powered@NLHPC: This research was partially supported by the
supercomputing infrastructure of the NLHPC (ECM-02).}
PJK was supported by the Deutsche Forschungsgemeinschaft Heisenberg
programme (grant No.\ KA 4825/1-1), and the Academy of Finland ReSoLVE
Centre of Excellence (grant No.\ 307411). Part of the simulations were
performed using the supercomputers hosted by CSC -- IT Center for
Science Ltd.\ in Espoo, Finland, who are administered by the Finnish
Ministry of Education. {J.S. acknowledges funding from the European Union's
Horizon 2020 research and innovation program under the Marie Sk{\l}odowska-Curie grant
No.\ 665667.}





\input{paper.bbl}





\bsp	
\label{lastpage}
\end{document}